\documentclass{article}

    \PassOptionsToPackage{numbers, compress}{natbib}

\usepackage[preprint]{neurips2021}

\usepackage[utf8]{inputenc} 
\usepackage[T1]{fontenc}    
\usepackage{hyperref}       
\usepackage{url}            
\usepackage{booktabs}       
\usepackage{amsfonts}       
\usepackage{nicefrac}       
\usepackage{microtype}      
\usepackage{xcolor}         
\usepackage{graphicx}
\usepackage{subfig}

\usepackage{amsmath}
\usepackage{amssymb}
\usepackage{amsthm}

\usepackage{amsmath}
\usepackage{kotex}

\usepackage{multirow}
\usepackage{algorithm}
\usepackage{algpseudocode}

\usepackage{graphicx}
\usepackage{dcolumn}
\usepackage{bm}

\title{Physics-aware Reduced-order Modeling of Transonic Flow via $\beta$-Variational Autoencoder}

\author{%
  Yu-Eop Kang\thanks{These authors contributed equally to this work}\\
  Department of Aerospace Engineering\\
  Seoul National University\\
  Republic of Korea, Seoul 08826 \\
  \texttt{kye72594@snu.ac.kr} 
  \And
  Sunwoong Yang$^*$\\
  Department of Aerospace Engineering\\
  Seoul National University\\
  Republic of Korea, Seoul 08826 \\
  \texttt{sunwoongy@gmail.com} 
  \And
  Kwanjung Yee\\
  Institute of Advanced Aerospace Technology\\
  Seoul National University\\
  Republic of Korea, Seoul 08826 \\
  \texttt{kjyee@snu.ac.kr}
}

\begin{document}

\maketitle

\begin{abstract}
Autoencoder-based reduced-order modeling (ROM) has recently attracted significant attention, owing to its ability to capture underlying nonlinear features. However, \textcolor{black}{two critical drawbacks severely undermine its scalability to various physical applications: entangled and therefore uninterpretable latent variables (LVs) and the blindfold determination of latent space dimension. In this regard, this study proposes the physics-aware ROM using only interpretable and information-intensive LVs extracted by $\beta$-variational autoencoder, which are referred to as physics-aware LVs throughout this paper. To extract these LVs, their independence and information intensity are quantitatively scrutinized in a two-dimensional transonic flow benchmark problem. Then, the physical meanings of the physics-aware LVs are thoroughly investigated and we confirmed that with appropriate hyperparameter $\beta$, they actually correspond to the generating factors of the training dataset, Mach number and angle of attack. To the best of the authors' knowledge, our work is the first to practically confirm that $\beta$-variational autoencoder can automatically extract the physical generating factors in the field of applied physics. Finally, physics-aware ROM, which utilizes only physics-aware LVs, is compared with conventional ROMs, and its validity and efficiency are successfully verified.}
\end{abstract}

\section{\label{introduction} Introduction}

\textcolor{black}{Computational fluid dynamics (CFD) has been widely applied in numerous engineering disciplines. However, high-fidelity CFD simulations are often computationally intensive due to the fine discretization of space and time domains. A practical approach to reduce its computational burden is using a regression model as a cost-efficient alternative to time-consuming flow analysis \cite{yang2022design}. Such regression models have been developed extensively, to name a few, cubic spline interpolation \cite{mckinley1998cubic}, the radial basis function \cite{gutmann2001radial}, and Gaussian process regression (GPR) \cite{rasmussen2003gaussian, schulz2018tutorial}; however, these models are only suitable for estimating scalar quantities such as lift coefficients of the airfoil.} Consequently, when the quantities of interest are high-dimensional vectors, such as pressure or the velocity field, the computational complexity becomes prohibitive owing to ``the curse of dimensionality.''

Data-driven reduced-order modeling (ROM) has recently attracted attention for its potential to deal with this problem. It treats a high-fidelity computer simulation as a ``black-box function'' and generates simplified models in a data-driven manner without any modification of the governing equation. The main purpose of this approach is to reduce the degrees of freedom of the data using the dimensionality reduction (DR) technique (also known as representation learning or manifold learning), which finds the low-dimensional latent representation of high-dimensional original data. Through this technique, the dimensionality of the data can be significantly reduced to a level that is suitable for training the regression models effectively.

One of the most widely used DR techniques is proper orthogonal decomposition (POD) \cite{sirovich1987turbulence}, which is also referred to as principal component analysis or the Karhunen-Lo\`eve theorem. Given the training data, POD extracts a set of orthogonal bases (also referred to as modes) that maximizes the variance of the projected data. \textcolor{black}{Extracted modes are ranked by their energy content so that the dimensionality can be reduced by truncating non-dominant modes. Specifically, the original high-dimensional data can be represented in a low-dimensional subspace using linear combinations of the dominant modes. However, its linearity makes POD-based ROM suffer from performance degradation in nonlinear problems \cite{amsallem2012nonlinear, lucia2002domain, kanglocal, dupuis2018surrogate}, requiring an excessive number of modes compared to nonlinear DR methods for the same reconstruction accuracy}.

\textcolor{black}{In this regard,} deep neural networks (DNNs) have become an alternative to POD for their performance on highly nonlinear problems. The most commonly used DNN-based DR technique is the autoencoder (AE), which learns the low-dimensional latent space of the original data in an unsupervised manner \cite{kramer1991nonlinear}. The nonlinear functions between its \textcolor{black}{multiple} hidden layers enable it to model \textcolor{black}{nonlinearity, and in this context Hinton et al. referred to AE as a nonlinear generalization of POD \cite{hinton2006reducing}. Owing to this property, numerous ROM studies have recently adopted AE as DR model for highly nonlinear problems \cite{milano2002neural, eivazi2020deep, zhang2021machine, hasegawa2020machine, wang2021flow, mucke2021reduced, wu2021reduced, maulik2021reduced, gruber2022comparison, kadeethum2022non}, and they confirmed that AE-based ROM shows higher reconstruction accuracy than the POD-based ROM in various nonlinear problems \cite{milano2002neural, eivazi2020deep, zhang2021machine}. Moreover, the flow field reconstruction using AE can be further improved by applying convolutional neural networks (CNNs), which were introduced to train grid-pattern data efficiently \cite{hasegawa2020machine, wang2021flow, mucke2021reduced, wu2021reduced, maulik2021reduced, gruber2022comparison, kadeethum2022non}.}

\textcolor{black}{However, there are several issues to be addressed in AE-based ROM. First, AE does not guarantee disentangled latent representation since its training algorithm only aims to reduce the reconstruction error without considering the regularity in latent space. Its irregularity allows multiple features to be entangled in a single latent variable (LV) \cite{higgins2016beta}, making it difficult to interpret the physical meanings of LVs. Second, since the model architecture of AE should be predetermined before training, its latent dimension need to be blindly selected large enough to contain the entire information (otherwise, a trial-and-error process should be performed). In POD, it is possible to truncate trivial LVs since it provides the energy contents of LVs (or modes). However, in AE, its algorithm does not treat LVs hierarchically according to their information intensity so that theoretically, the entire information is evenly distributed to each LV. Therefore, all dimensions blindly selected should be used, as truncation of even only one LV can severely compromise its reconstruction accuracy. As a result, the resultant redundancy of the latent dimension will degrade both the physical interpretability of LVs (too many LVs are entangled) and the efficiency of AE-based ROM (too many regression models should be trained).}

The introduction of a variational autoencoder (VAE) and its improved understanding has been providing a clue to \textcolor{black}{these problems.} While AE only minimizes reconstruction error, VAE also regularizes latent space by minimizing Kullback-Leibler (KL) divergence term \cite{kingma2013auto}. The balance between these two terms in VAE can be adjusted by the hyperparameter $\beta$, and the corresponding model is referred to as $\beta$-VAE \cite{higgins2016beta}. \textcolor{black}{It is known that by weighting the KL-divergence term in $\beta$-VAE, two notable effects can be achieved. First, it learns disentangled latent representations: a single LV encodes only one representation feature of the original dataset and therefore becomes interpretable. This ensures independence between each LV as the POD extracts orthogonal bases. Second, the regularization loss (KL-divergence) encourages learning the most efficient latent representation \cite{higgins2016beta, burda2015importance, burgess2018understanding}. Therefore, regardless of the predetermined latent dimension, only necessary LVs are automatically activated to contain the essential information \cite{eivazi2022towards}.} In summary, unlike AE model, $\beta$-VAE enables its latent space to be information-intensive without being entangled, indicating its potential to further improve the performance of AE-based ROM.

However, there are few studies related to $\beta$-VAE in the field of fluid dynamics. \textcolor{black}{Eivazi et al. \cite{eivazi2022towards} adopted $\beta$-VAE to extract interpretable nonlinear modes for time-dependent turbulence flows and proposed a method to rank LVs by measuring their energies. However}, their ranking method requires separate post-processing of the reconstructed data after model training, and the energies of LVs are indirectly calculated in the output space rather than in the latent space. Accordingly, the approach is unrelated to KL-divergence which is the main cause of information discrepancies between LVs. \textcolor{black}{Though they finally concluded that their framework successfully extracted interpretable features of turbulent flows, it lacks objectivity since it was based on the visual inspection of the flow fields. Last but not least, their research ended up with nonlinear mode decomposition without showing how interpretable LVs can be efficiently utilized for the ROM process. Indeed,} Wang et al. \cite{wang2021flow} already have applied $\beta$-VAE to ROM for transonic flow problems and verified its superior performance over POD. Though they extended $\beta$-VAE to practical application, they only focused on the implementation of $\beta$-VAE \textcolor{black}{for ROM so that the interpretability of extracted features (main purpose of $\beta$-VAE) and their effects on ROM performance were not addressed and referred to as their future work}.

\textcolor{black}{This study aims to utilize physically interpretable and information-intensive LVs obtained by $\beta$-VAE, which are referred to as ``physics-aware LVs'', for the efficient ROM process. For this purpose, a two-dimensional (2D) transonic flow problem is adopted as benchmark case, and the independence and information intensity of LVs are investigated first to confirm whether they are physics-aware. Herein, it was practically confirmed for the first time in the field of applied physics that $\beta$-VAE can automatically extract the physical generating factors. Then, ``physics-aware ROM'', ROM which utilizes only physics-aware LVs is proposed for its efficiency in that the number of required regression models can be reduced significantly. The presented physics-aware ROM is compared with conventional ROMs, and finally, we successfully verify its validity and efficiency}.

The rest of this paper is organized as follows. \textcolor{black}{In Sec. \ref{sec:DR}, the DNN-based DR techniques are described in detail, and in Sec. \ref{sec:ROM}, the physics-aware ROM is newly proposed. In Sec. \ref{sec:num_exp}, the setup of the numerical experiment is presented. In Sec. \ref{sec:result}, the process of extracting physics-aware LVs and the discovery of their actual physical meanings are described, and the effectiveness of physics-aware ROM based on them is investigated. And finally, in Sec. \ref{sec:conclustion}, the conclusion of this study and future work are presented.}

\section{\label{sec:DR} Dimensionality reduction techniques}

\subsection{\label{sec:AE} Autoencoder (AE)}
The AE model is a widely used deep learning-based DR technique. The \textcolor{black}{main objective of} the AE is to output exactly what is inputted. \textcolor{black}{Its} structure consists of two parts: an encoder and a decoder. The input of the AE, $\textbf{x}$, is entered into the encoder for the compression and exits as LVs, $\textbf{z}$. Then, $\textbf{z}$ enters the decoder and exits as reconstructed data $\tilde{\textbf{x}}$. The encoder and decoder consist of a multi-layer perceptron (MLP), which makes it possible to model nonlinearity in the reduction and reconstruction processes. Because the objective of training the AE model is to reconstruct $\tilde{\textbf{x}}$ similar to the original data $\textbf{x}$, the loss function is defined by Eq. \ref{eq:MSE}, \textcolor{black}{where $N$ denotes the number of data samples}. Although its loss function is defined by the mean square error (MSE) in this study, any other error metrics, such as the binary cross entropy between $\textbf{x}$ and $\tilde{\textbf{x}}$, can be used depending on the properties of the data. \textcolor{black}{Note that the adopted MSE is the mean operation over sample-wise, and the summation over element-wise (pixel-wise) directions.}
\begin{equation}\
\label{eq:MSE}
\mathcal{L}_{AE} = \frac{1}{N}\sum\limits_{i=1}^N{(\textbf{x}_i-\tilde{\textbf{x}}_i)^2},
\end{equation}

However, this AE model has a \textcolor{black}{obvious} limitation: there is no training algorithm \textcolor{black}{for guaranteeing} a regularized latent space. Herein, the expression ``regularized latent space'' means that the latent space is trained to be smooth and continuous; thus, when inputs $\textbf{x}_1$ and $\textbf{x}_2$ are similar (or close), their corresponding mapped latent space, $\textbf{z}_1$ and $\textbf{z}_2$, should also be similar \cite{ghosh2019variational}. \textcolor{black}{In AE model, a}s the original input $\textbf{x}$ becomes condensed layer-by-layer through the encoder, its representation becomes increasingly abstract \cite{vincent2008extracting}. Therefore, the output of the encoder $\textbf{z}$ is not guaranteed to be regularized in the training procedure of the AE. This explains why the \textcolor{black}{decoder part of the AE model cannot be used as a generative model: the trained latent space is irregular and therefore its correlation with the reconstructed data is abstract.}

\subsection{\label{sec:VAE} Variational autoencoder (VAE)}

Many studies have focused on the limitations of this unregularized latent space trained by an AE and have \textcolor{black}{alternatively} applied VAE in their framework. The structure of the VAE is similar to that of the AE. One major difference is that VAE stochastically extracts the latent space $\textbf{z}$ via random sampling, whereas the AE model obtains its latent space deterministically. 

The mathematical formulae for the VAE model are presented below (for further details, \textcolor{black}{please refer to these references \cite{kingma2013auto, yang2022inverse})}. We consider reducing the original dataset $\textbf{X} = \left\{\textbf{x}_{i} \right\}_{N}^{i=1}$ (assumed to be independently and identically distributed) to the latent space $\textbf{z}$ using the VAE model. In the inference of $\textbf{z}$ from $\textbf{X}$, variational inference is adopted due to the intractability of the posterior $p_\mathbf{\theta}(\textbf{z}|\textbf{x})$, where $\mathbf{\theta}$ is a parameter of the VAE model (to be more specific, it is intractable owing to the likelihood of $p_\mathbf{\theta}(\textbf{z})$). Therefore, instead of the intractable posterior $p_\mathbf{\theta}(\textbf{z}|\textbf{x})$, the VAE is trained to obtain its substitute, $q_\mathbf{\phi}(\textbf{z}|\textbf{x})$ ($\mathbf{\phi}$ is a variational parameter). Then, the log-likelihood of $p_\mathbf{\theta}(\textbf{x})$ can be expressed as:

\begin{equation}\
\label{eq:loglike}
\log(p_\mathbf{\theta}(\textbf{x}))=\int{\log{(\frac{p_\mathbf{\theta}(\textbf{x},\textbf{z})}{q_\mathbf{\phi}(\textbf{z}|\textbf{x})})}q_\mathbf{\phi}(\textbf{z}|\textbf{x})d\textbf{z}}+
\int{\log{(\frac{q_\mathbf{\phi}(\textbf{z}|\textbf{x})}{p_\mathbf{\theta}(\textbf{z}|\textbf{x})})}q_\mathbf{\phi}(\textbf{z}|\textbf{x})d\textbf{z}},
\end{equation}
\textcolor{black}{where} the second term on the right-hand side (RHS) is the KL-divergence of $q_\mathbf{\phi}(\textbf{z}|\textbf{x})$ from $p_\mathbf{\theta}(\textbf{z}|\textbf{x})$, \textcolor{black}{$KL(q_\mathbf{\phi}(\textbf{z}|\textbf{x})||p_\mathbf{\theta}(\textbf{z}|\textbf{x}))$}, which is always non-negative according to its definition (KL-divergence is a measurement of the statistical distance between two probability distributions). Therefore, the first term on the RHS becomes the lower bound of the log-likelihood \textcolor{black}{and} the problem of maximizing the log-likelihood becomes the problem of maximizing the lower bound. This lower bound can be expressed as:

\begin{equation}\
\label{eq:ELBO}
\int{\log{(\frac{p_\mathbf{\theta}(\textbf{x},\textbf{z})}{q_\mathbf{\phi}(\textbf{z}|\textbf{x})})}q_\mathbf{\phi}(\textbf{z}|\textbf{x})d\textbf{z}}=
\mathbb{E}_{q_\mathbf{\phi}(\textbf{z}|\textbf{x})}{[\log{p_\mathbf{\theta}(\textbf{x}|\textbf{z})}]}-
\int{\log{(\frac{q_\mathbf{\phi}(\textbf{z}|\textbf{x})}{p_\mathbf{\theta}(\textbf{z})})}q_\mathbf{\phi}(\textbf{z}|\textbf{x})d\textbf{z}},
\end{equation}
\textcolor{black}{where} the first and second terms on the RHS are the reconstruction error and KL-divergence of $q_\mathbf{\phi}(\textbf{z}|\textbf{x})$ from $p_\mathbf{\theta}(\textbf{z})$, respectively. However, owing to the existence of $q_\mathbf{\phi}(\textbf{z}|\textbf{x})$ in the reconstruction error, the back-propagation process cannot be performed, and calculating the gradient of the reconstruction error with respect to $\mathbf{\phi}$ is problematic \textcolor{black}{due} to the posterior $q_\mathbf{\phi}(\textbf{z}|\textbf{x})$. Therefore, a ``reparameterization trick'' is adopted, which allows for back-propagation \textcolor{black}{during} the sampling process. The concept behind this trick is to sample $\textbf{z}$ from the auxiliary noise variable $\epsilon \sim N(0,1^2)$: this random sampling makes the latent space in the VAE stochastically determined. To be more specific, the $\rm k^{th}$ LV ($z_k$) is assumed to follow the distribution below:
\begin{equation}\
\label{eq:latent_code}
z_k = \mu_k + \sigma_k \odot \epsilon.
\end{equation}
where $\odot$ denotes Hadamard product (element-wise product). Accordingly, the KL-divergence, the second term on the RHS in Eq. \ref{eq:ELBO} can be rewritten as below when the posterior $q_\mathbf{\phi}(\textbf{z}|\textbf{x})$ and prior $p_\mathbf{\theta}(\textbf{z})$ are assumed to follow the Gaussian distribution $N(\mathbf{\mu},\mathbf{\sigma}^2)$ and $N(0,\textbf{I}^2)$, respectively.
\begin{equation}\
\label{eq:KLwithrepar}
KL(q_\mathbf{\phi}(\textbf{z}|\textbf{x})||p_\mathbf{\theta}(\textbf{z})) = \frac{1}{2}\sum\limits_{k=1}^d{(\sigma_k^2+\mu_k^2-(\log(\sigma_k^2)+1))},
\end{equation}
\textcolor{black}{where} $\mu_k$ and $\sigma_k$ represent the mean and standard deviation used during the reparameterization of $z_k$, and $d$ denotes the dimension of the latent space. Herein, as the posterior approximation $q_\mathbf{\phi}(\textbf{z}|\textbf{x})$ approaches the prior $p_\mathbf{\theta}(\textbf{z})$, the KL-divergence decreases. Finally, the loss function of VAE model can be formulated as in Eq. \ref{eq:VAEloss}, which consists of the reconstruction error (MSE term) and regularizer (KL-divergence term). Since the KL-divergence term serves to regularize the latent space to be trained, it is also called regularization loss. It induces a sparser latent space \cite{kingma2013auto, doersch2016tutorial, higgins2016beta, pati2021attribute} just as the L1 regularization term makes the model sparse in the Lasso regression \cite{tibshirani1996regression}.
\begin{equation}\
\label{eq:VAEloss}
\begin{aligned}
\mathcal{L}_{VAE} 
&=\mathcal{L}_{AE} + KL(q_\mathbf{\phi}(\textbf{z}|\textbf{x})||p_\mathbf{\theta}(\textbf{z}))\\
&=\frac{1}{N}\sum\limits_{i=1}^N{(\textbf{x}_i-\tilde{\textbf{x}}_i)^2} + \frac{1}{2}\sum\limits_{k=1}^d{(\sigma_k^2+\mu_k^2-(\log(\sigma_k^2)+1))}.
\end{aligned}
\end{equation}

\subsection{\label{sec:bVAE} $\beta$-variational autoencoder ($\beta$-VAE)}

Higgins et al. \cite{higgins2016beta} focused on the trade-off relationship between the reconstruction error and KL-divergence in the loss function of the VAE. As the reconstruction accuracy increases, the degree of regularization (or information capacity) in the latent space decreases. To tune the balance between \textcolor{black}{these} two performances, Higgins et al. \cite{higgins2016beta} proposed the $\beta$-VAE, which can control the relative importance of the KL-divergence term using the adjustable hyperparameter $\beta$. It is a simple modification of the VAE: they have exactly the same structures but slightly different loss functions. The loss function of the $\beta$-VAE is as follows:
\begin{equation}\
\label{eq:bvae_loss}
\mathcal{L}_{\beta{\text -}VAE} =\frac{1}{N}\sum\limits_{i=1}^N{(\textbf{x}_i-\tilde{\textbf{x}}_i)^2}
+\beta KL(q_\mathbf{\phi}(\textbf{z}|\textbf{x})||p_\mathbf{\theta}(\textbf{z})). 
\end{equation}
The only difference in the loss functions between the VAE and $\beta$-VAE is whether the KL-divergence term is weighted by hyperparameter $\beta$. By introducing this hyperparameter, Higgins et al. \cite{higgins2016beta} achieved quantitative and qualitative improvements in the disentanglement within the latent representations over the traditional VAE model. \textcolor{black}{Finally, they concluded that owing to the disentangling performance of $\beta$-VAE, the interpretable representations of the independent generating factors of the given dataset can be discovered automatically. In this paper, the terminologies ``disentanglement'', ``interpretability'', ``independence'', and ``orthogonality'' are used interchangeably to refer to the following property of LVs: single LV stands for a single representation feature without the intervention of other LVs. However, what makes $\beta$-VAE special is not only in its disentangling performance. The regularization loss (KL-divergence) is known to have sparsification effect to encourage the most efficient representation learning \cite{higgins2016beta}. Burda et al. and Sønderby et al. practically confirmed this effect in that some LVs become inactive during the training process of VAE \cite{burda2015importance, sonderby2016ladder}. In particular, Burda et al. confirmed that inactive LVs have a negligible effect on the reconstruction. This finding indicates that VAE trains its LVs in an efficient manner by selectively activating LVs to contain the only necessary information. Since $\beta$-VAE model also has regularization term in the loss function and even weighted by the hyperparameter $\beta$, it can be easily inferred that the sparsification effect in the $\beta$-VAE model will become more dominant as $\beta$ increases. Taking these two outstanding advantages of $\beta$-VAE, disentangled and information-intensive latent space, a novel ROM framework is proposed in the next section.}

\section{\label{sec:ROM}{Physics-aware reduced-order modeling}}

\begin{figure}[htb!]
	\centering
		\includegraphics[width=1.\textwidth]{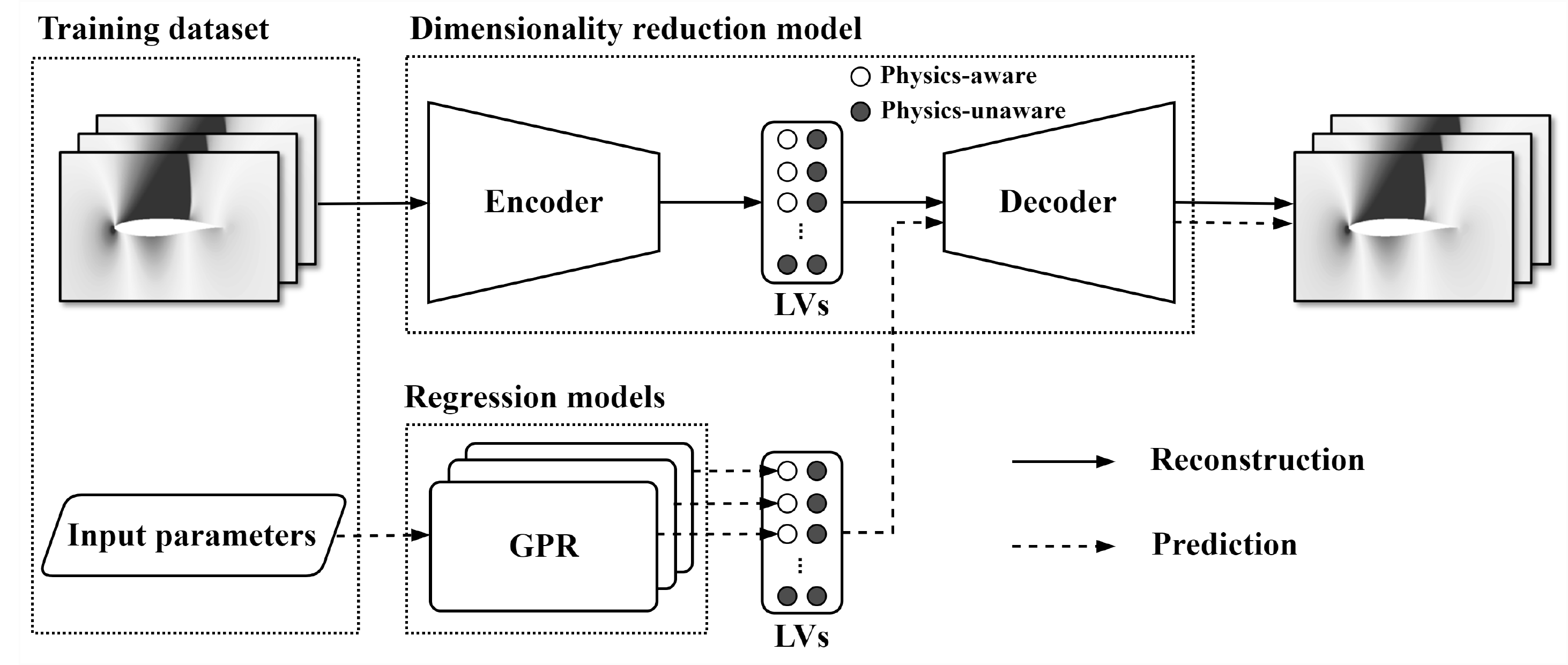}
	\caption{Overall structure of physics-aware reduced-order modeling.}
	\label{fig:ROM_structure}
\end{figure}

\textcolor{black}{The main goal of ROM is to predict the high-dimensional data with low computational cost. As described in Sec. \ref{sec:DR}, the high-dimensional data can be effectively reconstructed by the low-dimensional LVs through DR techniques. In this context, ROM aims to predict high-dimensional data efficiently by predicting these LVs from input parameters using regression models. Overall structure of ROM is illustrated in Fig. \ref{fig:ROM_structure}. In the reconstruction process (solid arrows), the high-dimensional data is encoded into LVs and reconstructed back into high-dimensional data by the decoder. In the prediction process (dashed arrows), LVs are first predicted from input parameters through regression models (GPR \cite{rasmussen2003gaussian, schulz2018tutorial} is adopted for regression in this study), and then predicted LVs are decoded into high-dimensional data. Since the high-dimensional data can be repeatedly predicted from input parameters at a very low computational cost, this prediction process is often referred to as the online phase.} 

\begin{figure}[htb!]
    \centering
	    \includegraphics[width=.7\textwidth]{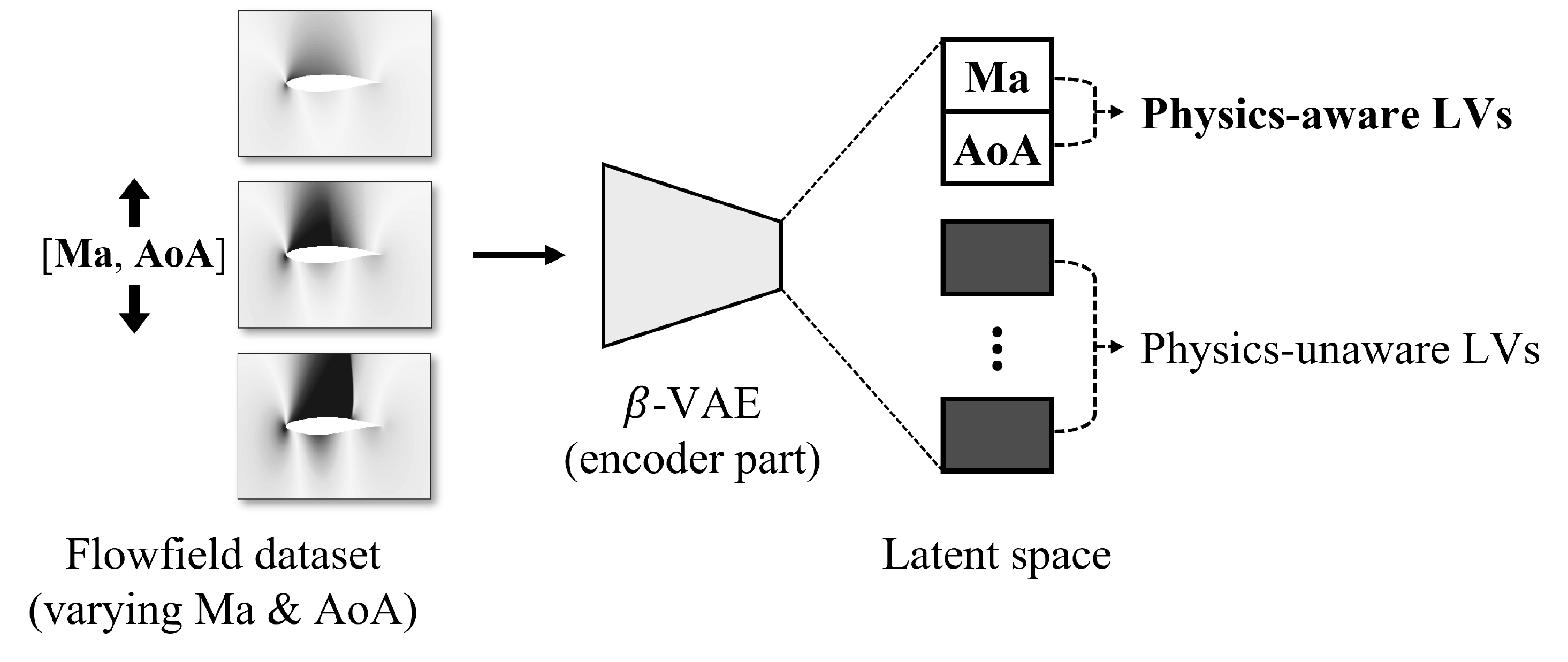}
	\caption{Illustrative schematic showing the process of extracting physics-aware LVs by $\beta$-VAE: the ideal case is to extract the actual physical parameters ($Ma$ and $AoA$) from the given dataset.}
	\label{fig:doe_to_disen}
\end{figure}

\textcolor{black}{Since LVs act as intermediaries in ROM for predicting high-dimensional data, they inevitably affect ROM performance. In this regard, AE model has two critical drawbacks to be applied to ROM. First, it trains the entangled and therefore uninterpretable latent space. Second, its latent dimension need to be blindly selected large enough since the model architecture should be predetermined before the training. Therefore, the AE-based ROM becomes physically uninterpretable and inefficient due to the excessive number of entangled LVs and the regression models to be trained. This study newly proposes physics-aware ROM via $\beta$-VAE to deal with these issues, considering the following characteristics of $\beta$-VAE. As stated in Sec. \ref{sec:bVAE}, the latent space in $\beta$-VAE is trained to be disentangled and information-intensive. When LVs actually satisfy these two properties simultaneously, they can be regarded as the latent representations which contain both interpretable and necessary information. In that this study uses physical dataset, LVs will correspondingly contain physical information, and accordingly, we refer to those LVs as ``physics-aware LVs.'' To ease the understanding of physics-aware LVs, the ideal schematic of their extraction with $\beta$-VAE is shown in Fig. \ref{fig:doe_to_disen}: when the dataset is generated through Mach number ($Ma$) and angle of attack ($AoA$), the ideally extracted physics-aware LVs will be $Ma$ and $AoA$. Finally, the ROM only utilizing these physics-aware LVs (which refers to ``physics-aware ROM'') is proposed for its efficiency that the number of regression models required in the prediction process can be significantly reduced.}

\textcolor{black}{In order to extract physics-aware LVs, we first estimate the independence of LVs. In this regard, Eivazi et al.\cite{eivazi2022towards} have already measured it using a Pearson correlation matrix and the same approach is applied herein. Then, the information intensity of LVs is quantified. Similar attempt has been made by Eivazi et al.\cite{eivazi2022towards}, who proposed a strategy to rank LVs by energy percentage, which quantifies the contribution of each LV to the reconstruction quality. However, they ranked LVs not in the latent space which is directly related to them, but in the output space so that there exist two problems. First, its low practicality due to cumbersome post-processing: flow fields should be reconstructed using a forward selection of LVs and then energy percentage is calculated from them. Second, since this method is irrelevant to KL-divergence, which is the main cause of the inactiveness of LVs \cite{burda2015importance, sonderby2016ladder}, it cannot be regarded as a fundamental approach for ranking LVs based on their amount of information. Therefore, another ranking criterion should be proposed that estimates the information intensity of each LV without such burdensome post-processing. For this purpose, we propose to apply the KL-divergence (Eq. \ref{eq:KLwithrepar}), the immediate cause for the sparser latent space due to its regularization effect. Since the calculation of KL-divergence is performed directly in the latent space, the decoder part of $\beta$-VAE does not even need to be utilized (no reconstruction required), solving all the limitations of the previous ranking approach.}

\textcolor{black}{In physics-aware ROM framework, only physics-aware LVs are utilized so that the number of regression models required in the prediction process can be significantly reduced. For example, suppose that AE-based ROM and $\beta$-VAE-based ROM are performed where both AE and $\beta$-VAE has the latent dimension of 16. In the case of AE-based ROM, 16 regression models should be trained because exclusion of just one LV can degrade ROM performance significantly in that all 16 LVs contain information in an entangled manner. However, in $\beta$-VAE-based ROM, since physics-unaware LVs are judged not to contain any meaningful information, it is sufficient to utilize only the regression models of physics-aware LVs. The overall procedure of physics-aware ROM is summarized in Algorithm \ref{alg:pa_rom}.}

\begin{algorithm}[htb!]
\caption{Physics-aware ROM via $\beta$-VAE}\label{alg:pa_rom}
\begin{algorithmic}

\State \textcolor{black}{(1) Prepare the training dataset generated by physical parameters.}
\State \textcolor{black}{(2) Train $\beta$-VAE with dataset in (1).}
\State \textcolor{black}{(3) Extract physics-aware LVs through estimating their independence by correlation coefficient and information intensity by KL-divergence.}
\State \textcolor{black}{(4) Train regression models to predict physics-aware LVs from physical parameters in (1).}
\State \textcolor{black}{(5) Predict the high-dimensional data from physical parameters using regression models trained in (4) and decoder part of $\beta$-VAE trained in (2). During this process, physics-unaware LVs are deactivated by being fixed to their estimated mean values.}

\end{algorithmic}
\end{algorithm}

\section{\label{sec:num_exp} Numerical experiments}

\subsection{\label{sec:problem_def} Data preparation}

\begin{figure}[htb!]
    \centering
    \includegraphics[width=0.65\linewidth]{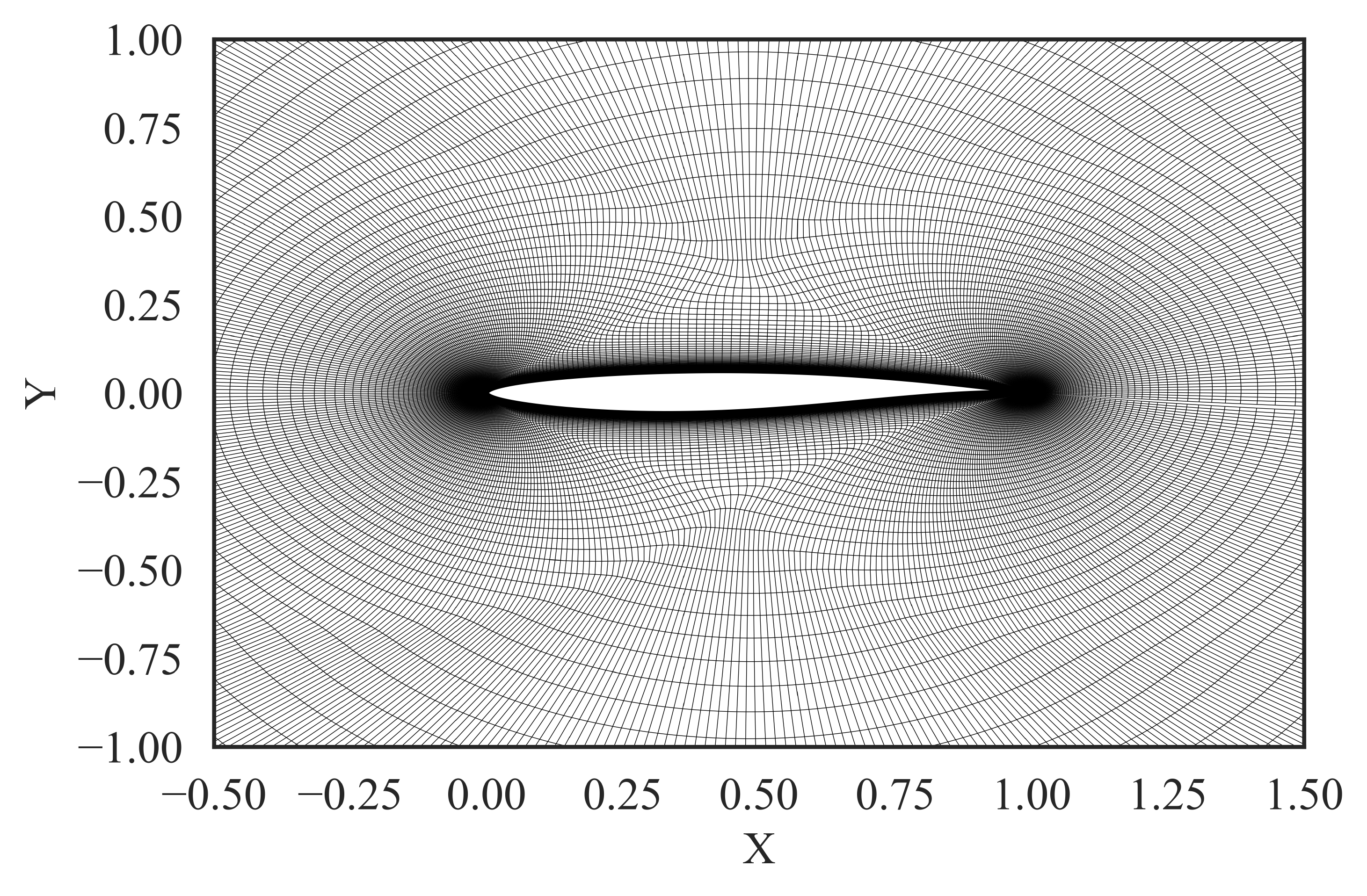}
    \caption{Computational grid used for the flow analysis; structured O-grid with a size of $512\times256$.}
    \label{fig:airfoil_grid}
\end{figure}

A 2D transonic flow problem, the benchmark case that is widely used in previous ROM studies \cite{kanglocal, dupuis2018surrogate, lucia2002domain, wang2021flow}, is adopted for validation of the proposed physics-aware ROM. The transonic flow field is generated by the KFLOW finite-volume-based CFD solver \cite{park2004implementation, hong2022enhanced}. Specifically, the Reynolds-averaged Navier-Stokes equation is solved coupled with the Spalart-Allmaras turbulence model. RAE 2822 airfoil is selected as the baseline geometry, and a structured O-grid with a \textcolor{black}{shape} of $512\times256$ \textcolor{black}{(wall-tangential direction $\times$ wall-normal direction)} is generated; the corresponding grid is shown in Fig. \ref{fig:airfoil_grid}. The Reynolds number is fixed at $6.5\times 10^6$, and two physical parameters are chosen: $Ma$ and $AoA$. The \textcolor{black}{design} space of each variable is set to [0.5, 0.8] and [$0^{\circ}$, $3^{\circ}$], respectively. Latin hypercube design is used to generate 500 sample points in 2D parameter space. A flow analysis \textcolor{black}{of these samples} is then conducted and the resultant velocity and pressure \textcolor{black}{fields} are normalized by \textcolor{black}{their} far-field conditions. As the variations of flow properties far from the airfoil are negligible, only the inner half of the entire grid \textcolor{black}{with respect to the normal direction from the airfoil is used so that the resultant grid becomes $512\times128$}. Consequently, a total dataset with a shape of $500 \times 3 \times 512 \times 128$ (dataset size $\times$ flow field components $\times$ \textcolor{black}{wall-tangential} direction grids $\times$ \textcolor{black}{wall-normal} direction grids) is obtained, where the flow-field components are the x-velocity, y-velocity, and pressure. Finally, the dataset size of 500 is split into a ratio of 9:1 \textcolor{black}{so that} the size of the training dataset is 450, and that of the test dataset is 50.

\subsection{\label{sec:param_set_lr} Training details}

\begin{figure*}[htb!]
	\centering
		\includegraphics[width=1.\textwidth]{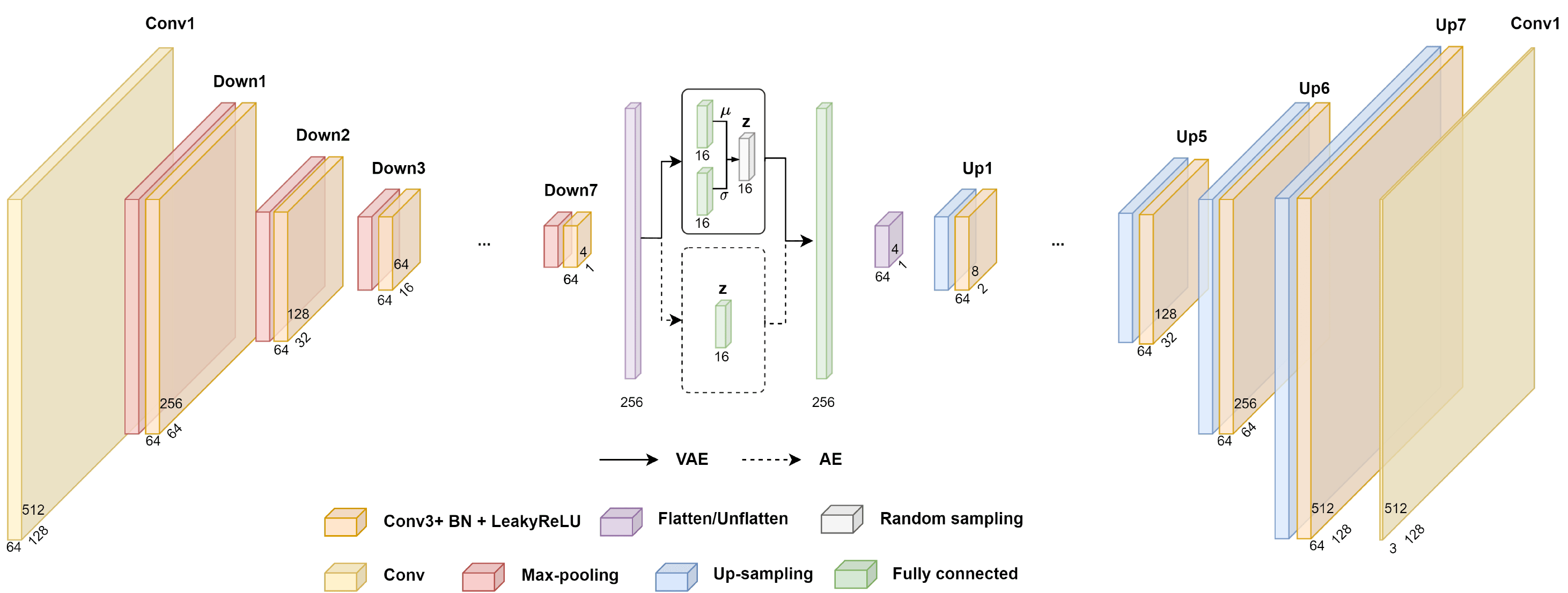}
	\caption{Structures of AE/VAE/$\beta$-VAE.}
	\label{fig:VAE_arch}
\end{figure*}

\begin{table}[htb!]

\caption{Details of the blocks and layers of AE/VAE/$\beta$-VAE used in this study.}\label{tab:NN_name}

\setlength{\tabcolsep}{0pt} 
\begin{tabular*}{\textwidth}{@{\extracolsep{\fill}\quad}lccccccccc}

\hline \hline
\textbf{Name}         & \textbf{Layer type}   & \textbf{Filter} & \textbf{Kernel} & \textbf{Stride} & \textbf{Activation} & \textbf{Batch Norm.} \\ \hline
\multirow{2}{*}{Up}   & Max-polling     & --               & 2 $\times$ 2           & --               &         --            & --                            \\
                      & Convolution     & 64              & 3 $\times$ 3           & 1               & LeakyReLU           & o                            \\\hline
\multirow{2}{*}{Down} & Upsampling      & --               & 2 $\times$ 2           & --               &         --           & --                            \\
                      & Convolution     & 64              & 3 $\times$ 3           & 1               & LeakyReLU           & o                            \\\hline
Conv1\_in             & Convolution     & 64              & 1 $\times$ 1           & 1               & LeakyReLU           & o                            \\\hline
Conv1\_out            & Convolution     & 3               & 1 $\times$ 1           & 1               & --                   & --                            \\\hline
FC                   & Fully Connected & --               & --               & --               & --                   & --                            \\   \hline \hline
\end{tabular*}
\end{table}

\begin{table}[htb!]
\caption{Network structure of the AE/VAE/$\beta$-VAE used in this study.}\label{tab:NN_structure}
\setlength{\tabcolsep}{0pt} 
\begin{tabular*}{\textwidth}{@{\extracolsep{\fill}}cccccc}

\hline \hline
\multicolumn{2}{c}{\textbf{Encoder}} & \multicolumn{2}{c}{\textbf{Bottleneck}} & \multicolumn{2}{c}{\textbf{Decoder}}  \\ \hline

\textbf{Layer} & \textbf{Output size} & \textbf{Layer} & \textbf{Output size} & \textbf{Layer} & \textbf{Output size} \\ \hline

Input          & 3 $\times$ 512 $\times$ 128        & Flatten        & 256           & Up1            & 64 $\times$ 8 $\times$ 2           \\
Conv1\_in      & 64 $\times$ 512 $\times$ 128       & FC1: $\mu$    & 16            & Up2            & 64 $\times$ 16 $\times$ 4          \\
Down1          & 64 $\times$ 256 $\times$ 64        & FC2: $\sigma$ & 16            & Up3            & 64 $\times$ 32 $\times$ 8          \\
Down2          & 64 $\times$ 128 $\times$ 32        & Resampling     & 16            & Up4            & 64 $\times$ 64 $\times$ 16         \\
Down3          & 64 $\times$ 64 $\times$ 16         & FC3            & 256           & Up5            & 64 $\times$ 128 $\times$ 32        \\
Down4          & 64 $\times$ 32 $\times$ 8          & Unflatten      & 64 $\times$ 4 $\times$ 1    & Up6            & 64 $\times$ 256 $\times$ 64        \\
Down5          & 64 $\times$ 16 $\times$ 4          & --             & --            & Up7            & 64 $\times$ 512 $\times$ 128       \\
Down6          & 64 $\times$ 8 $\times$ 2           & --             & --            & Conv1\_out     & 3 $\times$ 512 $\times$ 128        \\
Down7          & 64 $\times$ 4 $\times$ 1           & --             & --            & Output         & 3 $\times$ 512 $\times$ 128        \\  \hline \hline 

\end{tabular*}
\end{table}

In this study, all the models mentioned in Sec. \ref{sec:DR}, AE, VAE, and $\beta$-VAE, are trained to investigate their differences in terms of ROM for transonic flow. In particular, several $\beta$-VAE models are trained ($\beta\in[10, 20, 30, 40, 50, 100, 150, 200, 500, 750, 1000, 2000, 3000, 4000]$) to investigate the effects of the $\beta$ value (technically speaking, \textcolor{black}{$\beta$-VAE} can be regarded as the VAE when $\beta$ has a value of 1). \textcolor{black}{For all models, the dimension of the latent space should be determined blindly before the model training. The selected dimensions should be sufficient for encoding the training data generated from the 2D parameter domain ($Ma$ and $AoA$). In this regard, we adopted the approach suggested by Wang et al. which infers the latent dimension of $\beta$-VAE considering the accuracy of the POD with the same dimension \cite{wang2021flow} (their assumption was that POD requires much more dimensions than $\beta$-VAE for the equivalent reconstruction accuracy, which was also proved in their paper). Finally, dimension of the latent space is determined to be 16 since it conserves 99.18$\%$ in terms of energy contents of POD, which is judged to be sufficient.}

An appropriate encoder/\textcolor{black}{decoder} structure should be selected to effectively \textcolor{black}{compress/reconstruct} the data, \textcolor{black}{from}  dimensions of 3$\times$512$\times$128 to 16 \textcolor{black}{and vice versa}. To determine suitable structures, hyperparameter tuning based on a grid search was conducted. Finally, it was confirmed that the MSE reconstruction error, which represents the overall accuracy of the trained model, does not strongly depend on whether batch normalization, max-pooling, and up-sampling are applied. However, their application significantly affects the smoothness of the reconstructed flow field \textcolor{black}{(it can be inferred that this is due to the effects of batch normalization and max-pooling that prevent overfitting, and the interpolation effect of up-sampling)}. If an artificial discontinuity (rather than a discontinuity that reflects a physical phenomenon, such as a shock wave) is observed in the reconstructed flow field, the flow field cannot be considered realistic. Finally, the architectures of the selected models, which are considered to be sufficient in terms of MSE error and the smoothness of the reconstructed flow field, are shown in Tables \ref{tab:NN_name} and \ref{tab:NN_structure} and Fig. \ref{fig:VAE_arch}. The only difference between the structure of the AE and VAE/$\beta$-VAE is the bottleneck, and the structures of the VAE and $\beta$-VAE are exactly the same. 

The Adam optimizer is adopted to train selected models. The initial learning rate is set to $10^{-3}$, and it decays at a rate of 0.1 every 1000 epochs. The maximum epoch should be selected carefully because a model that does not fully converge can make significantly different predictions from a converged model \cite{yang2022comment}. The maximum number of epochs is set to 3000 \textcolor{black}{since it is verified to} guarantee sufficient convergence in terms of the loss function. A mini-batch size of 50 is selecte\textcolor{black}{d so that} nine iterations are performed per epoch \textcolor{black}{(as the size of the training dataset is 450)}. Using a Tesla P100-PCIE-16GB GPU \textcolor{black}{with Pytorch deep learning library \cite{paszke2019pytorch}}, the average training time for each model is calculated to be approximately 3 h.

\section{\label{sec:result} Result and discussion}

\textcolor{black}{In this section, the results of physics-aware ROM and physics-unaware ROM are presented in the following order. First, training results of AE/VAE/$\beta$-VAE are demonstrated to investigate the flow reconstruction accuracy. Next, the methods to measure the independence and information intensity of LVs are presented to determine whether the LVs obtained by AE/VAE/$\beta$-VAE are physics-aware LVs. At the same time, the validity of the KL-divergence ranking method to measure the information intensity of LVs is confirmed both quantitatively and qualitatively. Then, it is thoroughly investigated with various techniques whether the physics-aware LVs actually have interpretable physical features. In the end, it is verified that the physics-aware ROM based on these physics-aware LVs has equivalent prediction accuracy much more efficiently than physics-unaware ROM.}

\subsection{\label{sec:reconstruction} Training results}
\begin{figure}[htb!]
	\centering
		\includegraphics[width=.65\textwidth]{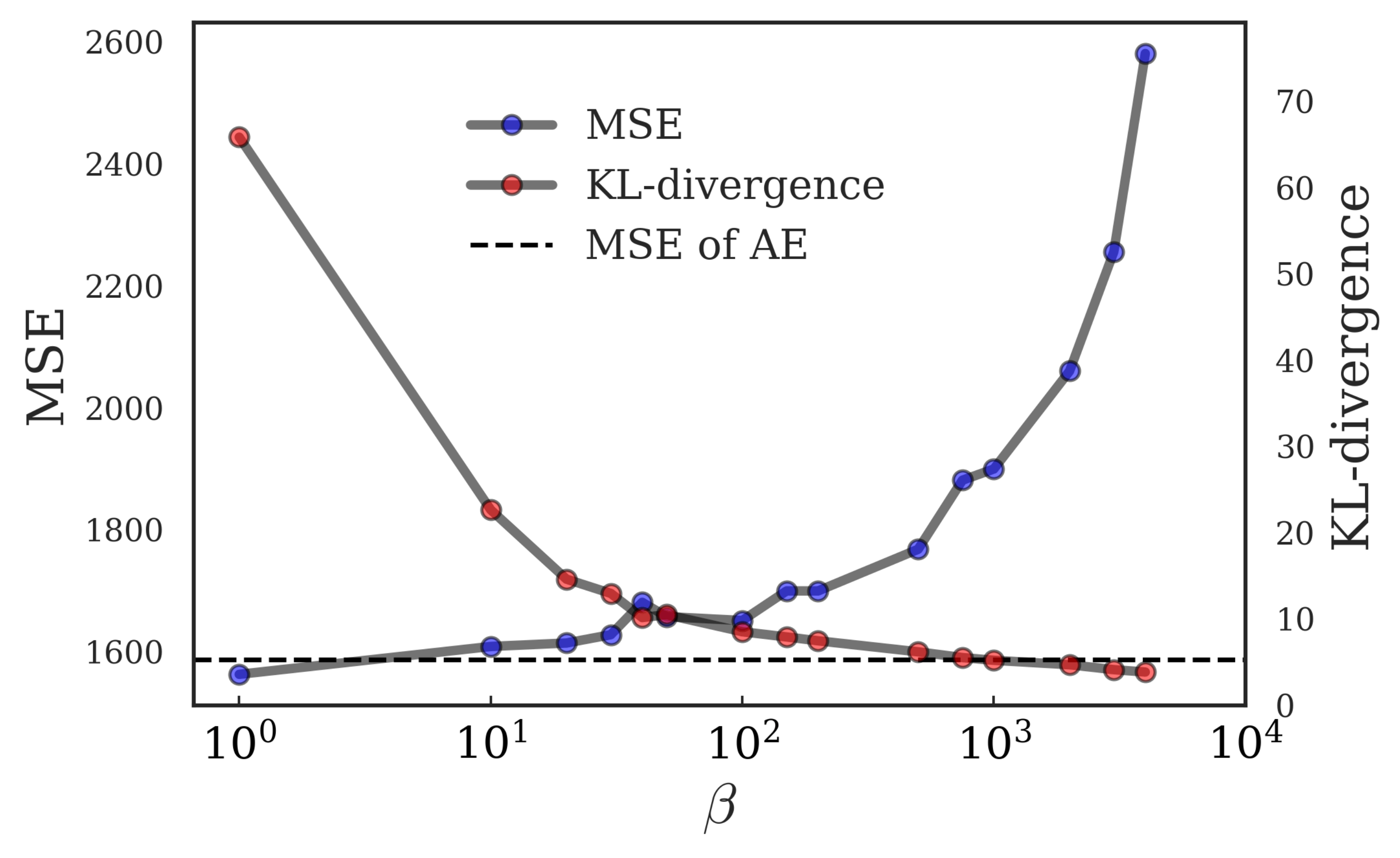}
	\caption{MSE and KL-divergence of the trained VAE/$\beta$-VAE models.}
	\label{fig:loss_tradeoff}
\end{figure}

First, the resultant loss functions of the \textcolor{black}{trained} models are shown in Fig. \ref{fig:loss_tradeoff}. More specifically, the loss function is decomposed into the MSE and KL-divergence (as in Eq. \ref{eq:bvae_loss}) to confirm their trade-off relationship. The MSE and KL-divergence of the VAE/$\beta$-VAE models are indicated by blue and red symbols, respectively. 
The $\beta$ and KL-divergence values cannot be defined in the AE model; therefore, only \textcolor{black}{its} MSE loss is indicated separately by a dashed line. Herein, a clear trade-off relationship between MSE and KL-divergence can be confirmed, as mentioned by Higgins et al. \cite{higgins2016beta}. As $\beta$ increases, KL-divergence decreases, while MSE increases. This is consistent with the concept of the $\beta$-VAE, which suppresses KL-divergence by increasing $\beta$. In particular, when $\beta>1000$, the MSE increases significantly \textcolor{black}{so that} an accurate reconstruction of the flow field is no longer possible.

\begin{figure*}[htb!]
	\centering
		\includegraphics[width=1.\textwidth]{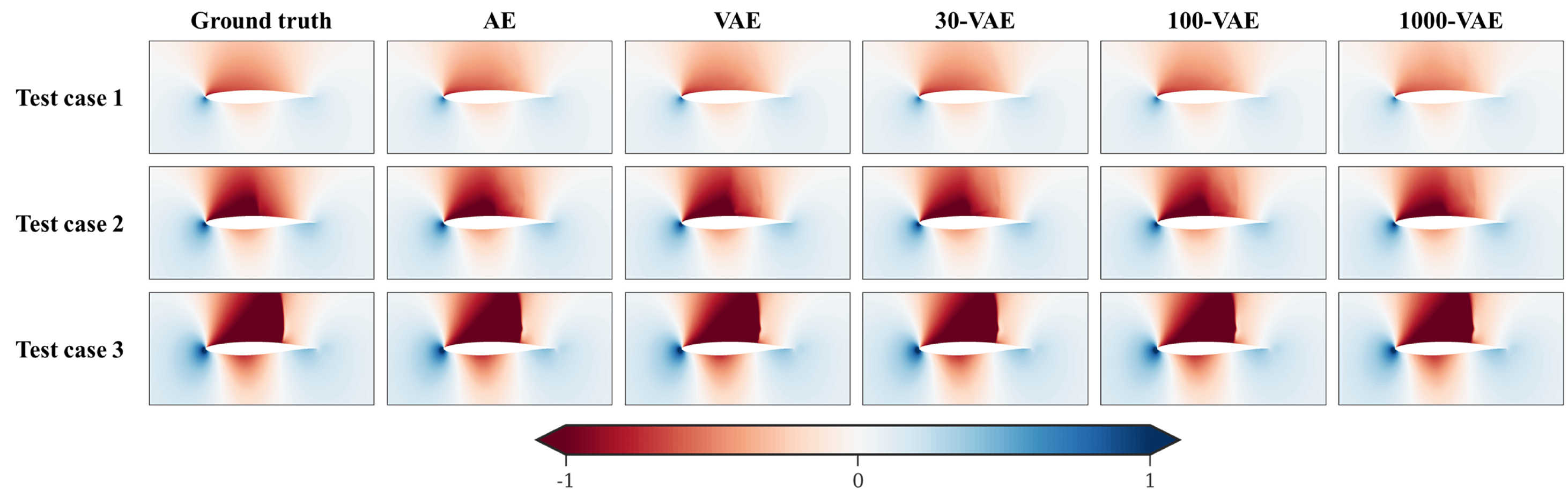}
	\caption{Reconstructed pressure fields of the trained models.}
	\label{fig:recon_compar}
\end{figure*}

Second, the reconstructed pressure flow fields of the three test cases (which are not used during the training process) are shown in Fig. \ref{fig:recon_compar} to compare the trained models more intuitively and visually. The selected test cases are as follows: test case 1 is in the absence of a shock wave ($Ma=0.61$ and $AoA=2.12^{\circ}$), test case 2 is in the presence of a weak shock wave ($Ma=0.72$ and $AoA=1.68^{\circ}$), and test case 3 is in the presence of a strong shock wave ($Ma=0.78$ and $AoA=1.57^{\circ}$). Five models are compared, including the AE, VAE, 30-VAE ($\beta$-VAE with $\beta=30$), 100-VAE  ($\beta=100$), and 1000-VAE ($\beta=1000$). In Fig. \ref{fig:recon_compar}, it can be confirmed that the reconstructed pressure contours for all three cases and all models do not exhibit any significant difference, indicat\textcolor{black}{ing} all models have been trained \textcolor{black}{to reconstruct the accurate} flow fields. 

\subsection{\label{sec:independence} Independence of LVs}

\begin{figure}[htb!]
    \centering
	    \includegraphics[width=.8\textwidth]{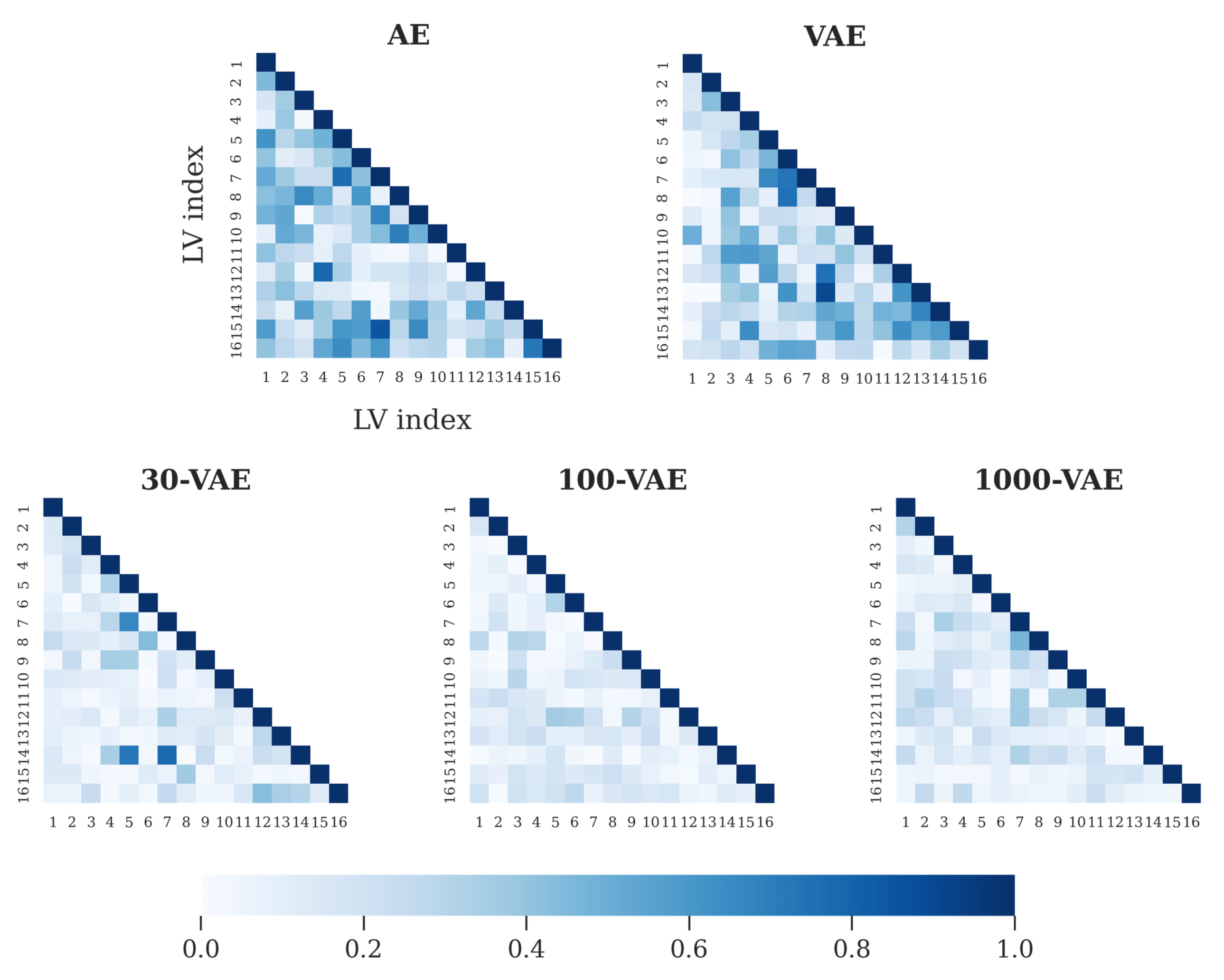}
	\caption{Absolute values of the components in the Pearson correlation matrices for LVs.}
	\label{fig:heatmap}
\end{figure}

\begin{figure}[htb!]
    \centering
	    \includegraphics[width=.9\textwidth]{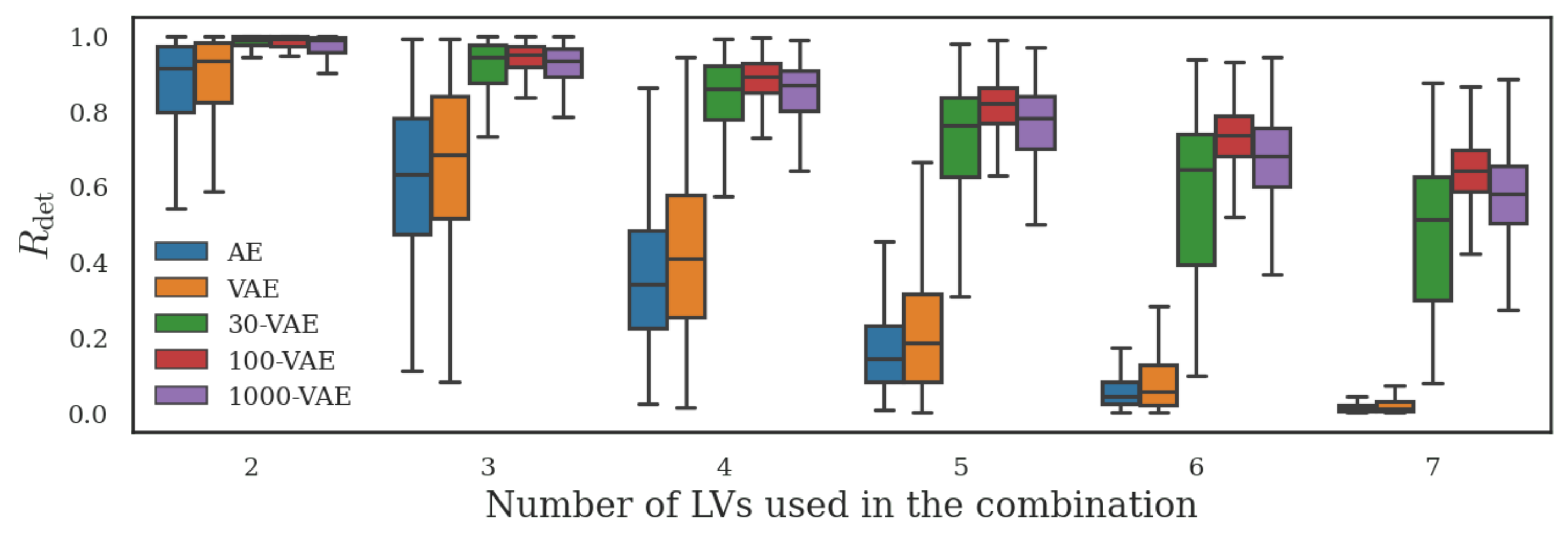}
	\caption{Determinants of Pearson correlation matrices for combinations of two to seven LVs.}
	\label{fig:combi}
\end{figure}

\textcolor{black}{This section is to confirm whether LVs of $\beta$-VAE are actually trained to be disentangled from each other so that they are interpretable.} In Fig. \ref{fig:heatmap}, the absolute values of the components in the Pearson correlation matrix are shown from AE to 1000-VAE. Since there is no algorithm in AE model to promote the independence of LVs, it has the largest values. However, in $\beta$-VAEs, their values decrease as $\beta$ increases, which indicates that the LVs gradually become independent of each other. These results practically prove that the KL-divergence actually encourages the independence of each LV. \textcolor{black}{To measure the degree of correlation within the entire set of LVs, Eivazi et al. computed the determinant of the correlation matrix (when the determinant is 0/1, it indicates that these variables are completely correlated/uncorrelated) \cite{eivazi2022towards}. However, given the fact that the KL-divergence term leads to a sparser latent space, which will be discussed in detail in Sec. \ref{sec:activness}, examining the determinant of the whole matrix size of $16 \times 16$ is contradictory.} Therefore, the independence of various LV combinations is further analyzed. First, all possible combinations of two to seven LVs are obtained. Then, the determinants of these combinations are calculated and their statistics are summarized in Fig. \ref{fig:combi}. For any number of LVs used in a combination, the $\beta$-VAE models have significantly higher determinants than the AE and VAE models. In summary, the results presented in Fig. \ref{fig:heatmap} and \ref{fig:combi} consistently show that the $\beta$-VAE successfully learns uncorrelated LVs compared to AE and VAE models, owing to the KL-divergence, which forces the LVs to be independent.

\subsection{\label{sec:activness} Information intensity of LVs}  

\begin{figure}[htb!]
    \centering
	    \includegraphics[width=.65\textwidth]{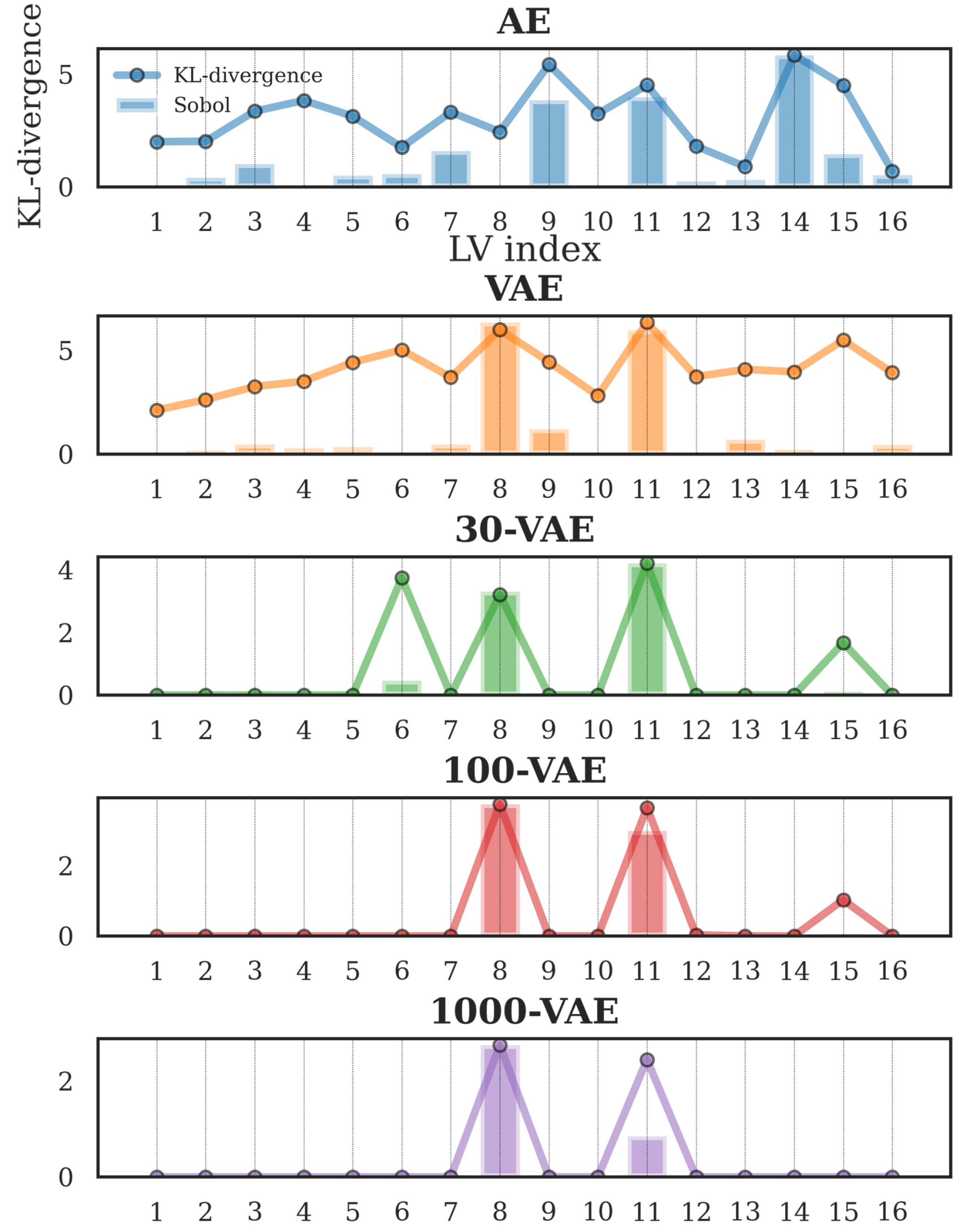}
	\caption{KL-divergence and Sobol results of each LV with respect to the training dataset.}
	\label{fig:KL_ANOVA}
\end{figure}

\textcolor{black}{It was verified in Sec. \ref{sec:independence} that the LVs in $\beta$-VAE are trained to contain disentangled information. In this section, we tried to rank such disentangled LVs according to their information intensity using KL-divergence. Fig. \ref{fig:KL_ANOVA} shows the KL-divergence of each LV in AE/VAE/$\beta$-VAE. It can be confirmed that both the AE and VAE models do not have any inactive LVs. On the contrary, notable trends are observed in 30-VAE, 100-VAE, and 1000-VAE models: each model has only four (LV index 6, 8, 11, and 15), three (index 8, 11, and 15), and two (index 8 and 11) activated LVs. These results are consistent with those of Eivazi et al. in that the number of activated LVs decreases as beta increases in the $\beta$-VAE model \cite{eivazi2022towards}.} In fact, a similar approach was applied by Sønderby et al., who judged whether the LV is activated via the KL-divergence \cite{sonderby2016ladder}. However, the relationship between the activeness of the LV and KL-divergence was not described clearly. Therefore, we also attempt to clarify it: further investigations are conducted to check whether the LV judged to be more active in terms of the KL-divergence actually has a greater effect on the reconstructed flow field. Sobol sensitivity analysis is utilized for this analysis \cite{saltelli2010variance,Herman2017}. A total 18,432 latent vectors are sampled using the Saltelli sampler in the range of $[\hat{\mu}_{k}-2\hat{\sigma}_{k}, \hat{\mu}_{k}+2\hat{\sigma}_{k}]$, where $\hat{\mu}_{k}$ and $\hat{\sigma}_{k}$ each represents estimated mean and standard deviation of $\rm k^{th}$ LV with respect to the training dataset. The corresponding reconstructed flow fields are obtained through the decoder parts of the AE/VAE/$\beta$-VAE models. Since Sobol analysis should be conducted on the scalar output, a set of 3$\times$512$\times$128 pixels of the flow fields is converted to a scalar value. The sum of all pixel components is used in this study, but any scalar value representing the main characteristics of the flow field can be used. Fig. \ref{fig:KL_ANOVA} shows the calculated first-order Sobol indices. It can be confirmed that the higher the KL-divergence, the larger the value of the Sobol indices in the VAE/$\beta$-VAE models (because \textcolor{black}{KL-divergence has nothing to do with the algorithm of the} AE model, it does not exhibit a clear trend with the Sobol indices). 

\begin{figure}[htb!]
    \centering
	    \includegraphics[width=.65\textwidth]{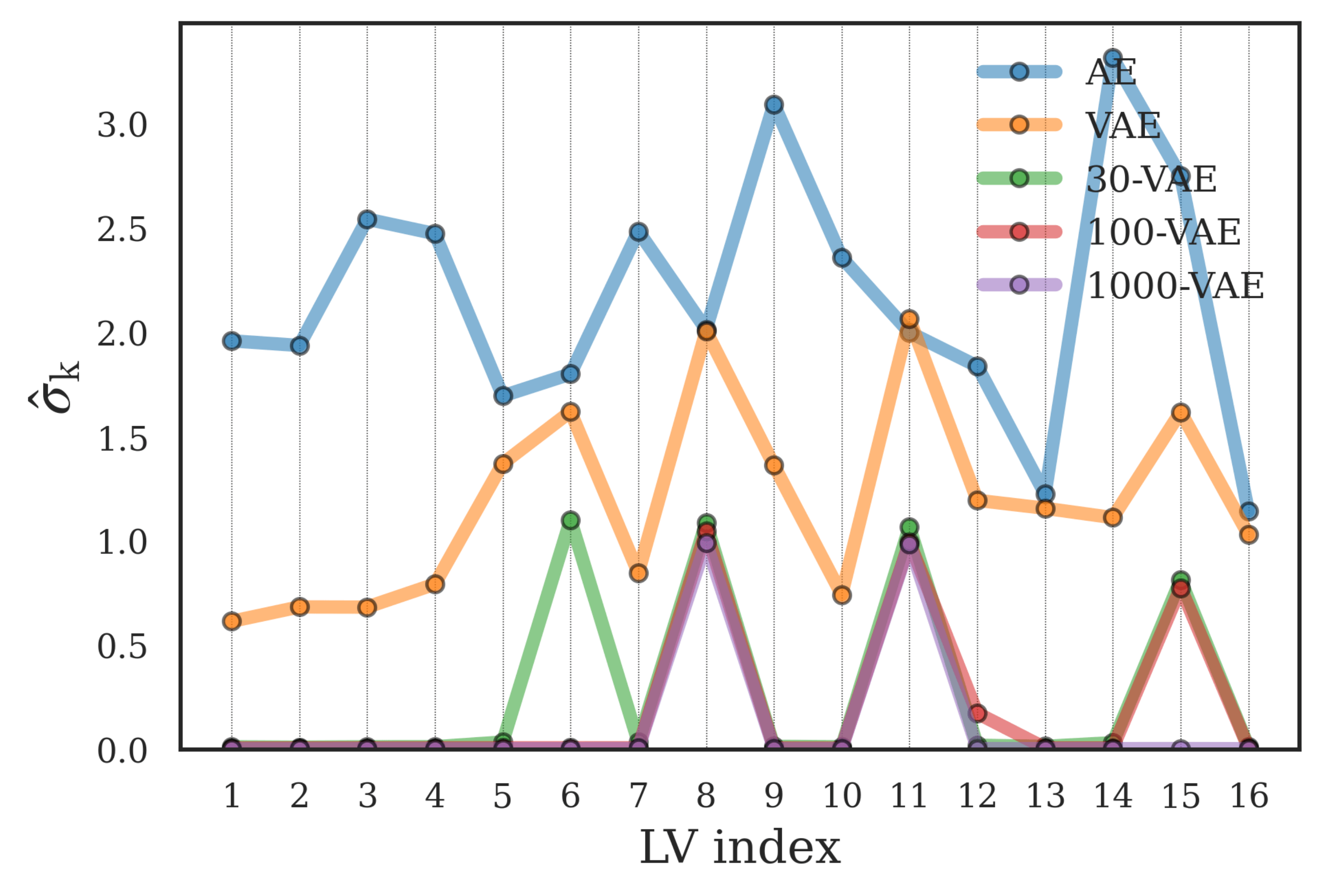}
	\caption{Estimated standard deviations of LVs with respect to the training dataset.}
	\label{fig:LV_std}
\end{figure}

\textcolor{black}{Additionally, since the standard deviation represents the dispersion of the variable, which can be regarded as its activeness}, $\hat{\sigma}_{k}$ values are also investigated. When $\hat{\sigma}_{k}$ is low for a specific LV, it can be understood that the corresponding LV remains inactive (or less dispersed) in the latent space during training: a situation where the value of the LV does not change even if the input data changes. In Fig. \ref{fig:LV_std}, it can be confirmed again that both AE and VAE models do not have any inactive LVs in that all $\hat{\sigma}_{k}$ values are larger than 0.5. However, the most interesting point is that compared to Fig. \ref{fig:KL_ANOVA}, LVs with high KL-divergence also have high $\hat{\sigma}_{k}$ in 30-VAE, 100-VAE, and 1000-VAE models: the LV indices of the activated LVs in terms of KL-divergence exactly match those of $\hat{\sigma}_{k}$. To sum up, we suggest the use of KL-divergence as a criterion for ranking the activeness (information intensity) of LVs for the two reasons. First, it is a regularization loss, which is the direct cause of the information discrepancies between LVs. Second, it does not require any cumbersome post-processing. The justification of this decision-making process is successfully performed through intuitive but quantitative investigations by comparing its ranking with the Sobol indices (Fig. \ref{fig:KL_ANOVA}) and estimated standard deviations from the training dataset (Fig. \ref{fig:LV_std}).

\begin{figure}[htb!]
    \centering
	    \includegraphics[width=1.\textwidth]{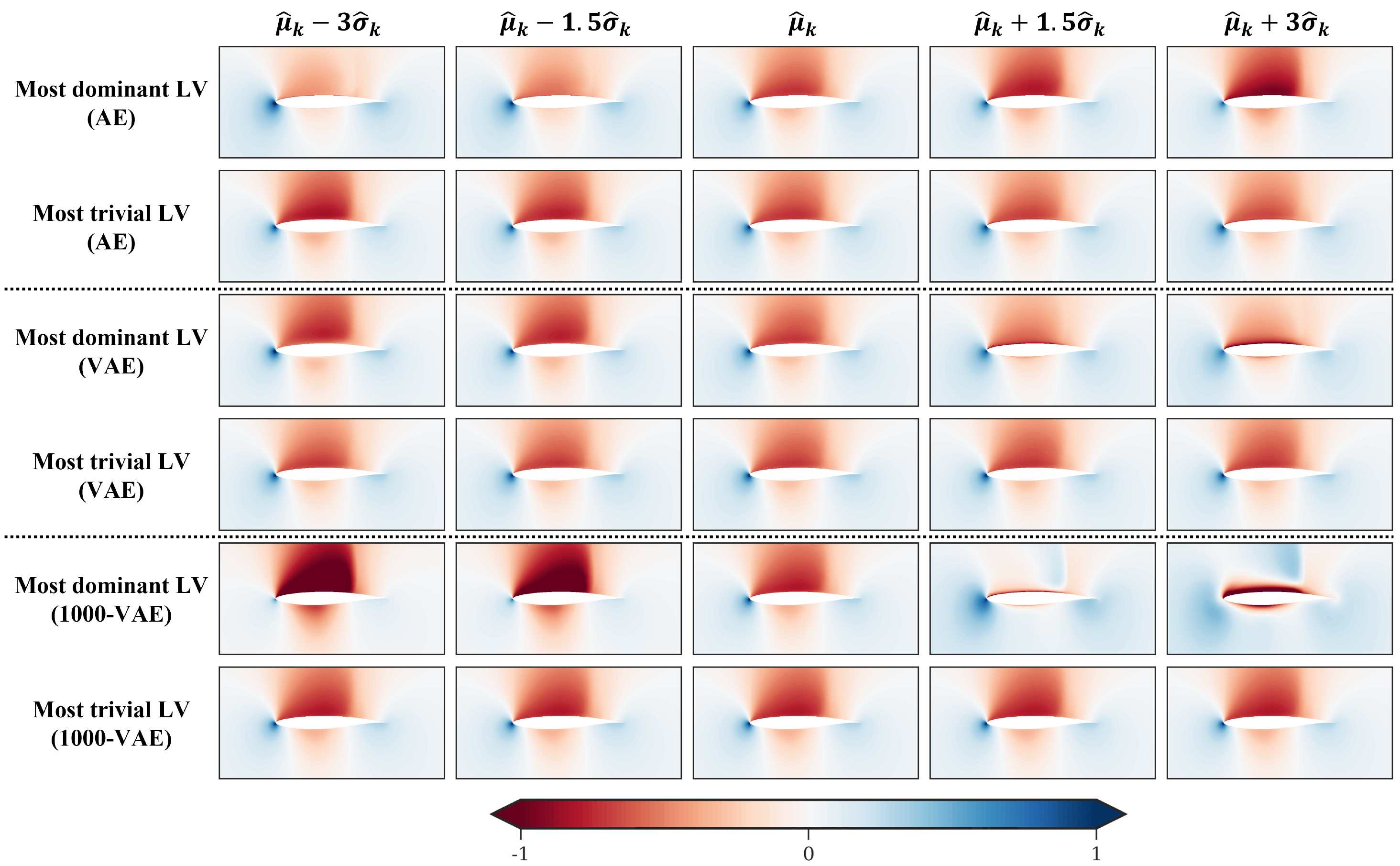}
	\caption{Latent traversal plots of pressure flow fields for two extreme LVs: first (most dominant) and last (most trivial) LVs ranked by KL-divergence.}
	\label{fig:lat_trav}
\end{figure}

So far, using proposed KL-divergence criterion, the inactiveness within the latent space has been investigated in a quantitative manner, but it does not provide straightforward information on what these activated/inactivated variables actually contain. \textcolor{black}{Accordingly, each LV is visualized via a traversal of itself, which is the most widely adopted approach for this purpose \cite{yang2022inverse, higgins2016beta}: it shows the traversal of a single LV while other LVs remain fixed so that one can visually understand the features of a specific LV without the intervention of other variables.} In this study, a latent traversal plot is applied to identify the physical features of the flow field contained in each LV. Fig. \ref{fig:lat_trav} shows the pressure flow fields for two extreme LVs: one is the most dominant LV, which is ranked first by KL-divergence, and the other is the most trivial LV, which is ranked last. In AE, the pressure field changes abruptly as the most dominant LV changes; when the most trivial LV changes, the field changes gradually, but not as significant as that of the most dominant LV. Since the sparsification effect does not occur in AE due to the absence of KL-divergence term in the loss function, all the variables are activated and therefore contain uninterpretable (or entangled) physical features. However, in VAE and 1000-VAE models, the most trivial LVs cause indistinguishable variations in the flow fields. This is because the KL-divergence forces inactiveness in the latent space, so only necessary LVs become activated to contain meaningful physical features: it can be regarded as a virtue of $\beta$-VAE model. When $\beta$ increases, information is packed into LVs more compactly \cite{eivazi2022towards}, as observed in the traversal of the most dominant LV in the 1000-VAE model. This LV can be interpreted as containing information on the occurrence of the shock wave, and the variation it arouses is the largest compared to other models.

\subsection{\label{sec:Disen} Physics-awareness of LVs}

\textcolor{black}{The two requirements for physics-aware LVs are investigated so far: whether LVs are disentangled (Sec. \ref{sec:independence}) or information-intensive (Sec. \ref{sec:activness}).} In this section, physics-oriented investigations are performed to figure out the physical meanings each physics-aware LV contains.

\begin{figure}[htbp!]

    \begin{minipage}[c]{0.3\textwidth}%
    \subfloat[\label{fig:ma_aoa_a}]{\includegraphics[clip,width=1\textwidth]{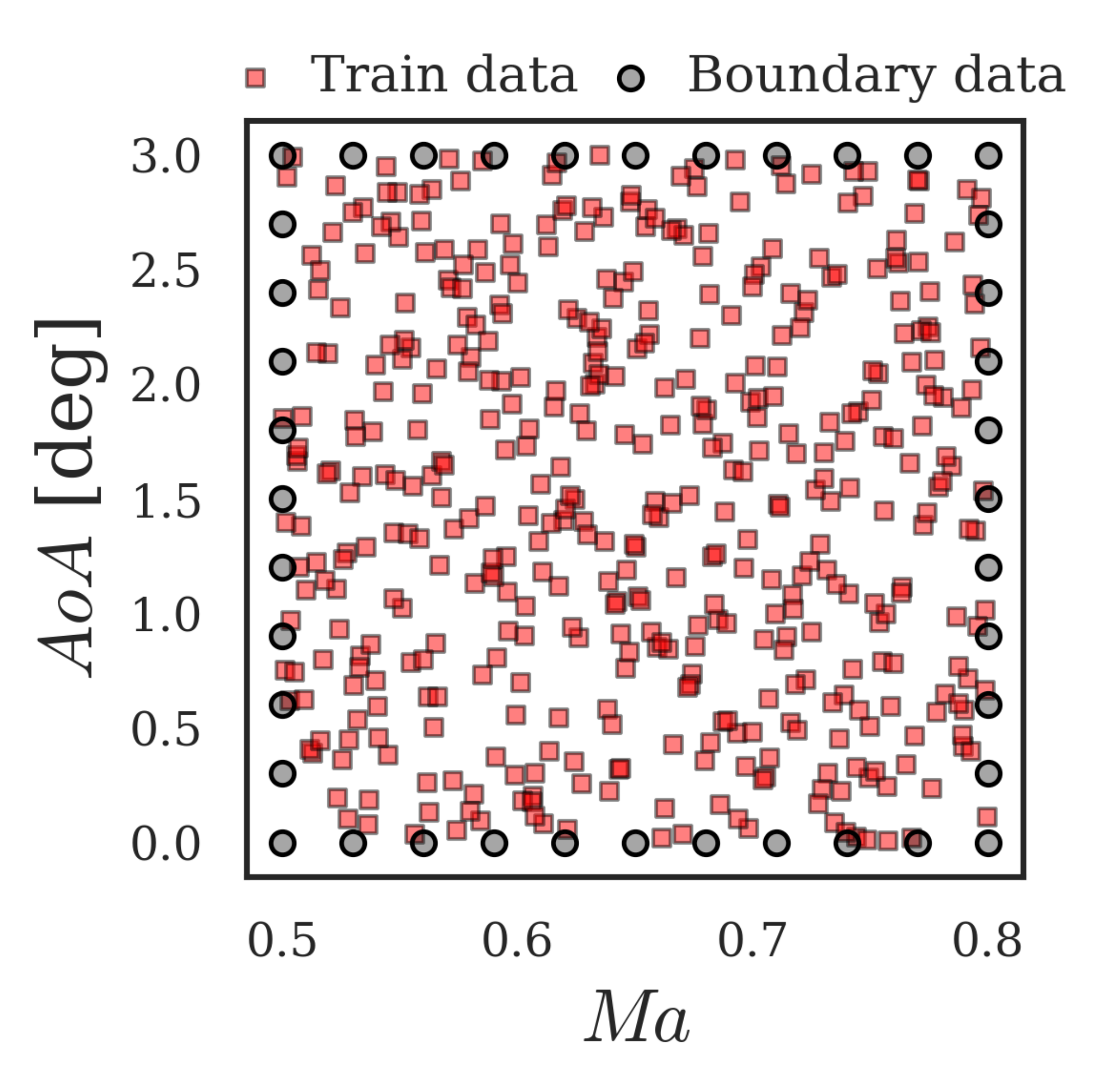}}%
    \end{minipage}
    \hfill
    \begin{minipage}[c]{0.7\textwidth}%
    \subfloat[\label{fig:ma_aoa_b}]{\includegraphics[clip,width=1\textwidth]{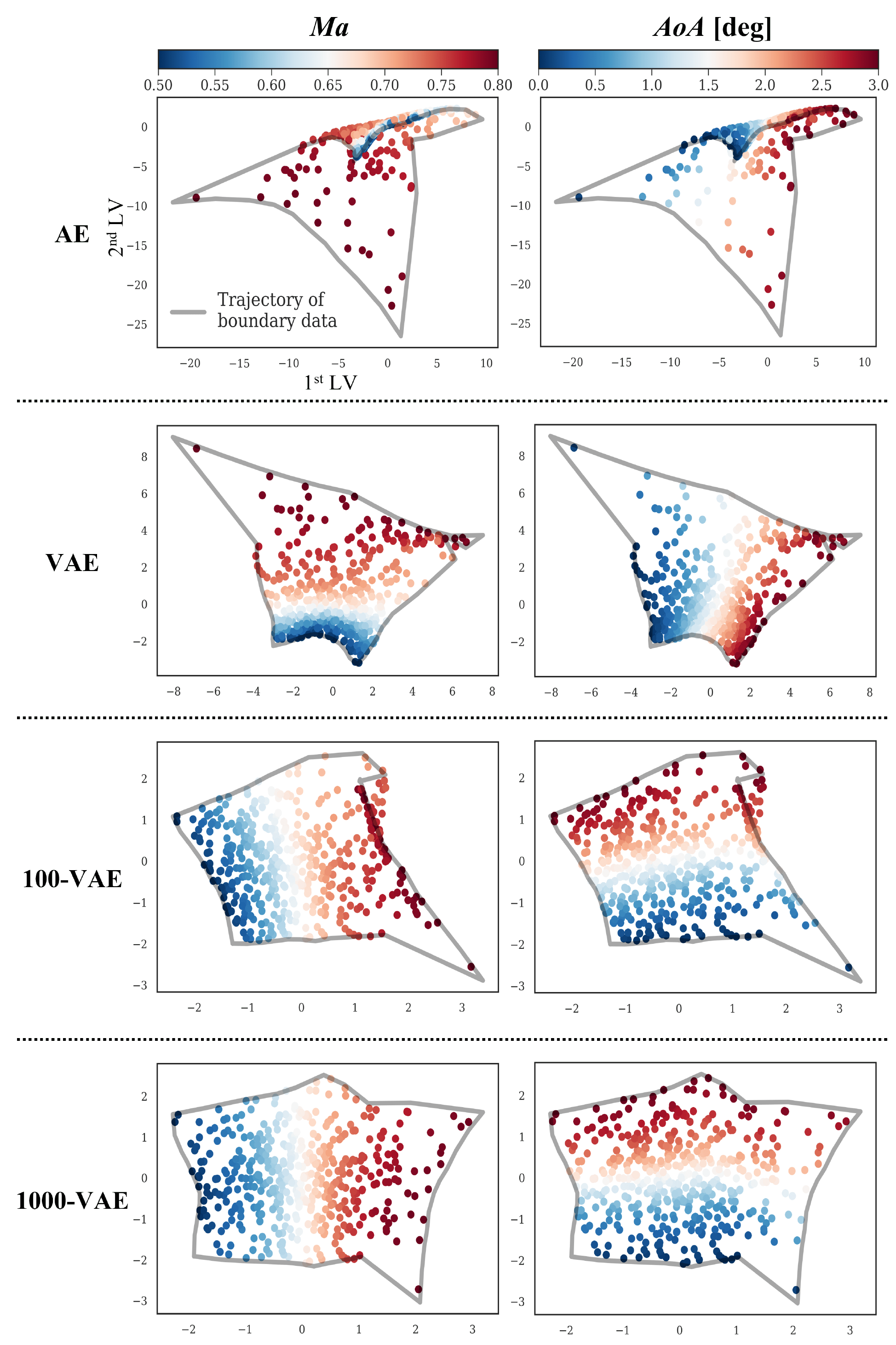}}%
    \end{minipage}
    
    \caption{Investigation of physical features contained in the top two LVs: (a) distributions of training dataset and boundary data with respect to $Ma$ and $AoA$, and (b) distributions of training dataset with respect to $1^\mathrm{st}$ and $2^\mathrm{nd}$ LVs (the left figures are colored by $Ma$, and the rights by $AoA$).}
     \label{fig:ma_aoa}
\end{figure}

\textcolor{black}{First, the relationship between the two physical parameters used in this study (whose distributions are shown in Fig. \ref{fig:ma_aoa_a}) and the top two ranked LVs by KL-divergence ($1^\mathrm{st}$ LV and $2^\mathrm{nd}$ LV) is visually investigated as $\beta$ varies. Accordingly, in Fig. \ref{fig:ma_aoa_b}, the distributions of the training dataset with respect to those two LVs are shown. For the AE, VAE, 100-VAE, and 1000-VAE models, the plots of the first column are colored by $Ma$ value, and the second column by $AoA$. Moreover, to confirm that how physical parameter domain can be represented by top two LVs, we also draw the trajectories of the boundary data in Fig. \ref{fig:ma_aoa_b} (they are extracted from the boundary of the physical parameter space, as in Fig. \ref{fig:ma_aoa_a}). In AE, neither $Ma$ nor $AoA$ have any noticeable trends. In contrast, in VAE, the plot in the first column exhibits a trend that is not clear, but still noticeable: as the $1^\mathrm{st}$/$2^\mathrm{nd}$ LV increases, $AoA$/$Ma$ increases. However, the increase in $Ma$ cannot be explained by the increase in the $2^\mathrm{nd}$ LV alone (the same goes for $1^\mathrm{st}$ LV and $AoA$). These ambiguous correlations between the two LVs and two physical parameters become more clear in 100-VAE (for this case, as the $1^\mathrm{st}$/$2^\mathrm{nd}$ LV increases, $Ma$/$AoA$ increases). Though 100-VAE shows much obvious relationship than AE, the physical parameters cannot be fully represented only by two dominant LVs in that its boundary trajectory does not cover the entire training dataset. Considering the fact that there are three physics-aware LVs in 100-VAE, this may be an expected result. However, it should be noted here that the top two variables sorted by KL-divergence almost succeeded in representing the physical parameters, whereas when the same plot is drawn with the 1st and 3rd dominant LVs, a very irregular pattern is observed. In this regard, the validity of ranking LVs through KL-divergence is confirmed once again. Finally, the 1000-VAE is investigated. Since this model has two physics-aware LVs, in the ideal case one can expect them to correspond to the two physical parameters (this ideal situation was depicted in Fig. \ref{fig:doe_to_disen}). And this conjecture is actually happening in 1000-VAE: the correlations between two physical parameters and two LVs become clear, and the latent space of the training dataset is perfectly closed by its boundary trajectory. Here is the key finding of this study: though $\beta$-VAE has no information about the physical parameters used to generate the training data, it effectively extracts only two LVs out of total 16 LVs (especially when $\beta$=1000), which correspond to actual physical parameters $Ma$ and $AoA$. It can be concluded that these marvellous results are owing to the orthogonality effect (makes LVs disentangled) and regularization effect (makes redundant LVs inactive) of the KL-divergence term in $\beta$-VAE. To make this clear, it should be noted that AE, one of the most widely used nonlinear DR techniques, fails to construct a physics-aware latent space in that it learns all 16 entangled LVs and therefore physically uninterpretable.}

\begin{figure}[htbp!]
	\centering
	\subfloat[\label{fig:lin_reg_a}]{\includegraphics[width=0.45\textwidth]{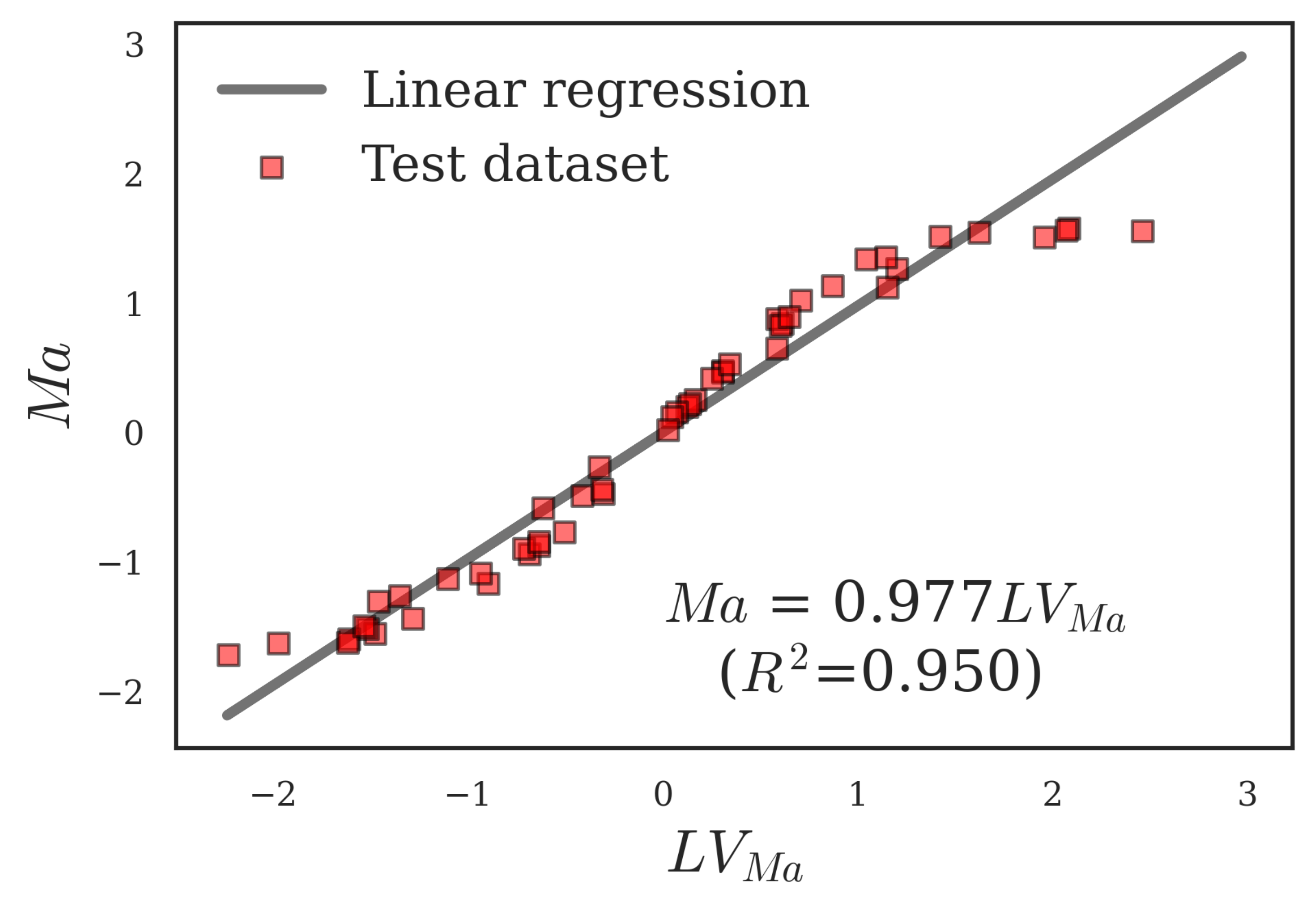}}
	\hfill
    \subfloat[\label{fig:lin_reg_b}]{\includegraphics[width=0.45\textwidth]{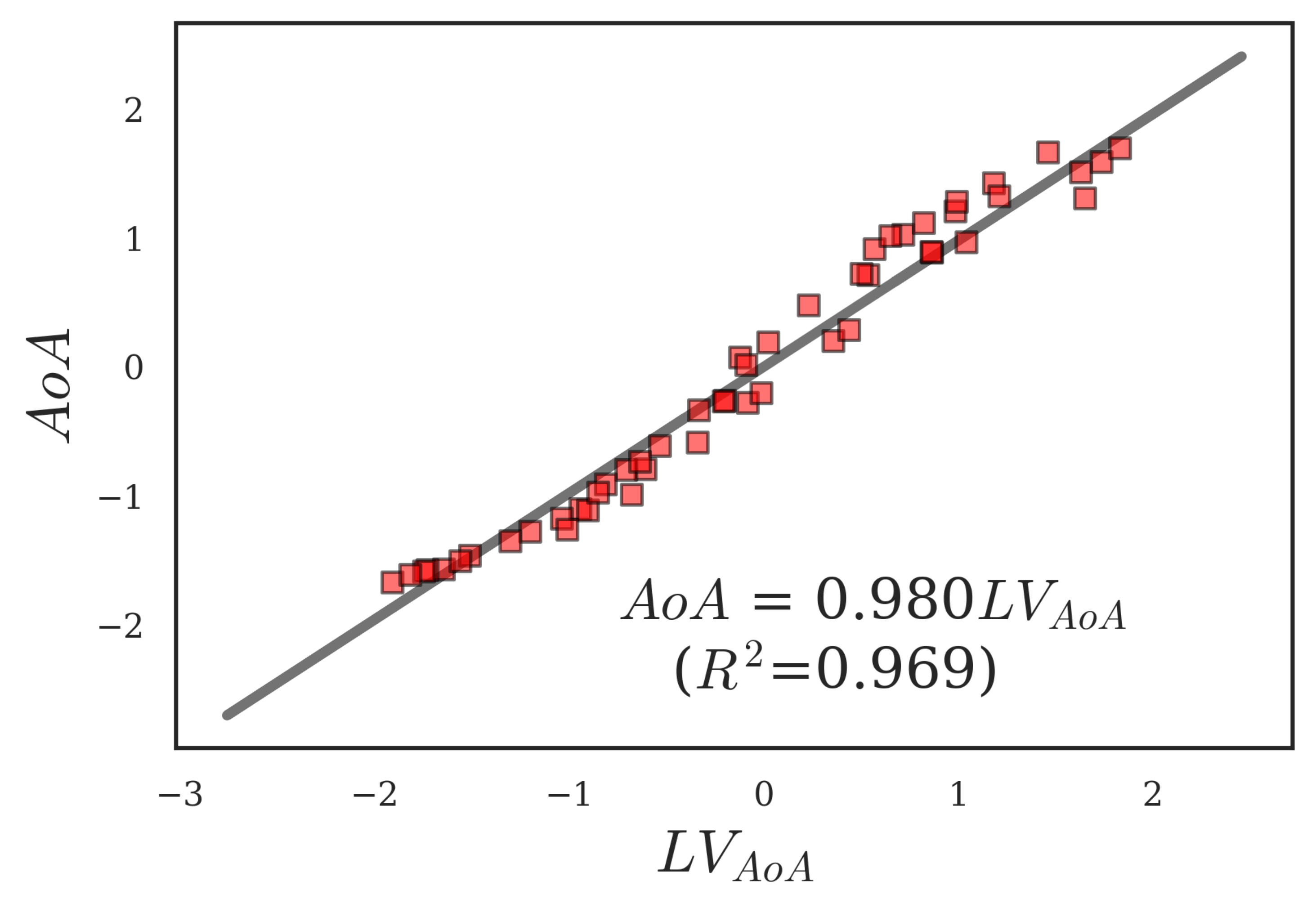}}

    \caption{The results of the single variable LR: (a) $Ma=f(LV_{Ma})$, and (b) $AoA=f(LV_{AoA})$.}
     \label{fig:lin_reg}
\end{figure}

As the visual analysis in the previous paragraph is qualitative, a quantitative analysis is performed herein to verify whether $LV_{Ma}$ or $LV_{AoA}$ in 1000-VAE corresponds to $Ma$ or $AoA$ (for the sake of brevity, $LV_{Ma}$ refers to the $1^\mathrm{st}$ LV, and $LV_{AoA}$ refers to the $2^\mathrm{nd}$ LV, which implies that $LV_{Ma}$ and $LV_{AoA}$ are the LVs responsible for $Ma$ and $AoA$, respectively). Single-variable linear regression (LR) models are trained \textcolor{black}{for this purpose}: since the relationships between physical parameters and LVs appear linear in Fig. \ref{fig:ma_aoa}, the LR model is considered to be sufficient. \textcolor{black}{For example}, LR model for $Ma$ is trained with the input variable as $LV_{Ma}$ and the output variable as $Ma$,which can be expressed as $Ma=f(LV_{Ma})$. The training dataset is the same as in Sec. \ref{sec:problem_def} and physical parameters and LVs are standardized before training. If $LV_{Ma}$ and $Ma$ (or $LV_{AoA}$ and $AoA$) have a linear relationship, the fitted LR models will perform well on the test dataset. The results are shown in Fig. \ref{fig:lin_reg}. In each subplot, the equation of the trained LR model is also included. \textcolor{black}{For both LR models \textcolor{black}{$Ma=f(LV_{Ma})$ and $AoA=f(LV_{AoA})$}, their equations clearly show that $Ma\approx{}LV_{Ma}$ and $AoA\approx{}LV_{AoA}$ in that the coefficients are approximately 1 and the intercepts are 0. Also, the coefficients of determination ($R^{2}$) calculated based on the test dataset are 0.950 and 0.969, respectively. It indicates that only one LV is sufficient to accurately represent each physical parameter.} In addition, we train two additional LR models with both LVs as input features: one \textcolor{black}{as $Ma=f(LV_{Ma},LV_{AoA})$ and the other as $AoA=f(LV_{Ma},LV_{AoA})$. If $Ma$ needs both LVs for its expression, the $R^{2}$ value of the LR model $Ma=f(LV_{Ma},LV_{AoA})$ will be significantly higher than that of $Ma=f(LV_{Ma})$.} The trained LR models are described as follows:
\begin{equation}\
\label{eq:bi_lr}
\begin{bmatrix}
Ma \\
AoA 
\end{bmatrix} = 
\begin{bmatrix}
0.978 & -0.011\\
-0.017 & 0.981
\end{bmatrix}
\begin{bmatrix}
{LV}_{Ma} \\
{LV}_{AoA}
\end{bmatrix}.
\end{equation}
Herein, there are two notable points. First, the coefficient of \textcolor{black}{$LV_{AoA}$ is negligible (approximately 0) compared to that of $LV_{Ma}$ (approximately 1) when modeling $Ma$, which indicates that $LV_{AoA}$ is redundant variable for representing $Ma$ (same principal applies when modeling $AoA$)}. The second interesting point is that the values of $R^{2}$ do not change compared to those of the single-variable LR models. \textcolor{black}{The $R^{2}$ of LR model $Ma=f(LV_{Ma},LV_{AoA})$ only increases 0.001 than $Ma=f(LV_{Ma})$, and $AoA=f(LV_{Ma},LV_{AoA})$ model has the same $R^{2}$ as $AoA=f(LV_{AoA})$. These two points quantitatively} suggest that the extracted physics-aware LVs actually correspond to $Ma$ and $AoA$ in a disentangled (or independent) manner.

\begin{figure}[htbp!]
	\centering
	\subfloat[\label{fig:cp_ma_aoa_a}]{\includegraphics[width=0.45\textwidth]{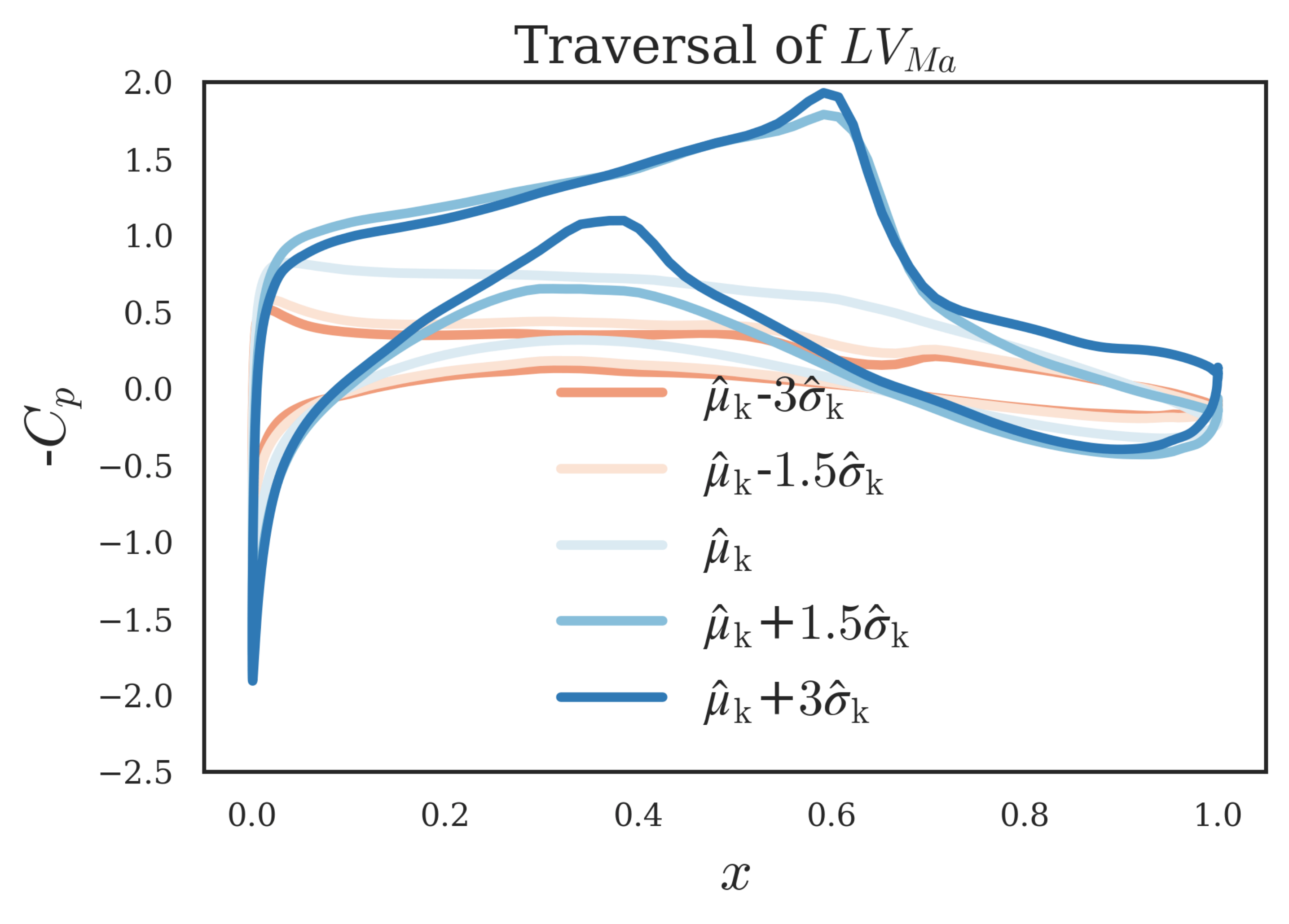}}
	\hfill
    \subfloat[\label{fig:cp_ma_aoa_b}]{\includegraphics[width=0.45\textwidth]{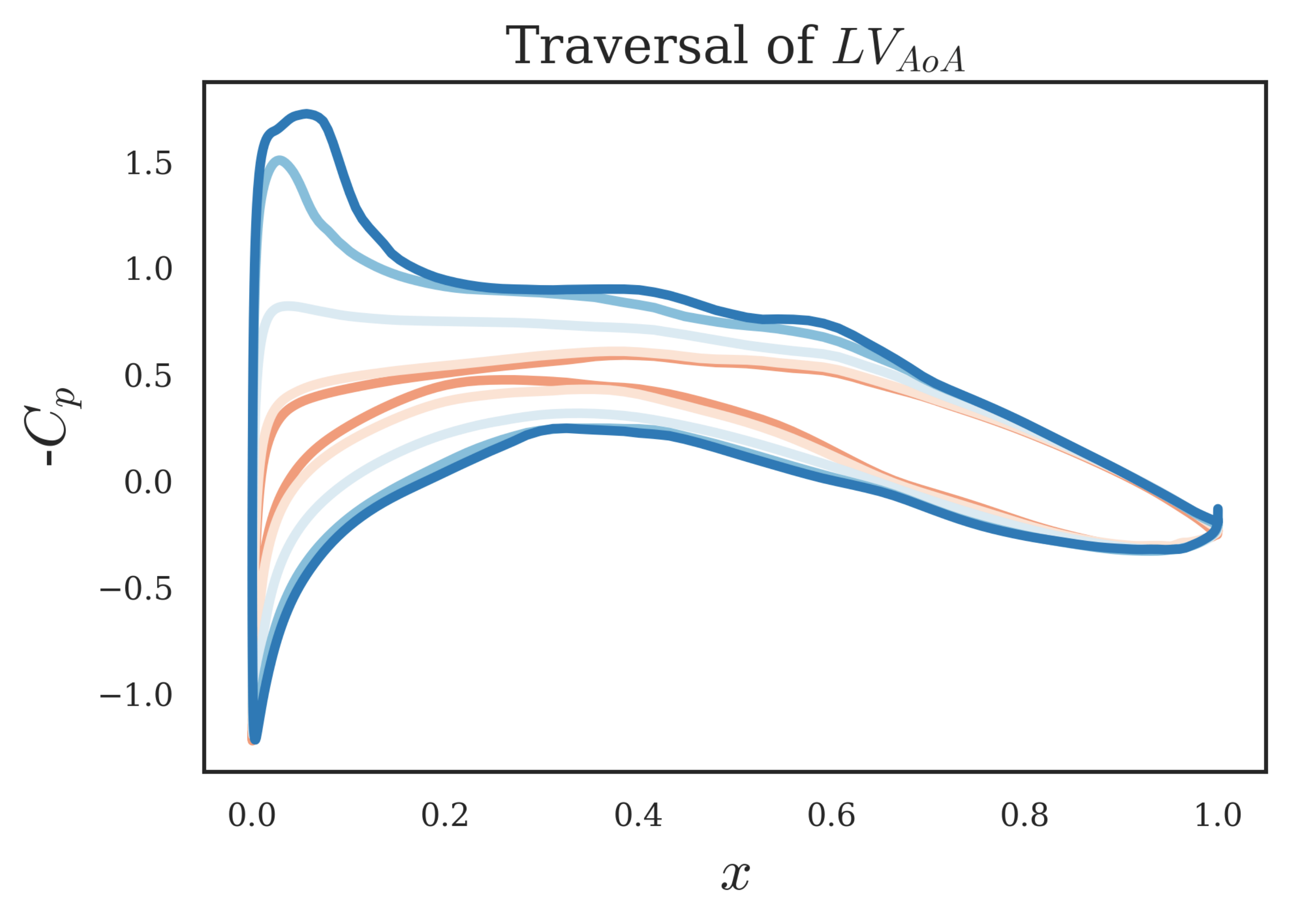}}

    \caption{Latent traversal plots of airfoil surface pressure distributions in 1000-VAE: (a) traversal of $LV_{Ma}$, and (b) traversal of $LV_{AoA}$.}
     \label{fig:cp_ma_aoa}
\end{figure}

\textcolor{black}{Since t}he physical meanings of $LV_{Ma}$ and $LV_{AoA}$ are verified both qualitatively and quantitatively, the latent traversal plots of airfoil surface pressure distributions are shown in Fig. \ref{fig:cp_ma_aoa} to intuitively check their impact on the flow field. In the traversal of $LV_{Ma}$ (Fig. \ref{fig:cp_ma_aoa_a}), when the LV is in the range of $\hat{\mu}_{k}-3\hat{\sigma}_{k}$ to $\hat{\mu}_{k}$, there is no shock wave, but as it increases to $\hat{\mu}_{k}+1.5\hat{\sigma}_{k}$ and $\hat{\mu}_{k}+3\hat{\sigma}_{k}$, the occurrence of a shock wave is shown. In the traversal of $LV_{AoA}$ (Fig. \ref{fig:cp_ma_aoa_b}), a variation in the location and magnitude of the leading edge suction peak is observed. This visual analysis of the actual effects of $LV_{Ma}$ and $LV_{AoA}$ on the pressure distributions also leads to the consistent conclusion that they correspond to $Ma$ and $AoA$, respectively. 

\textcolor{black}{Although the ability of $\beta$-VAE to extract physics-aware LVs is first observed with simple physical parameters in this study, it has tremendous potential to be utilized to extract generating factors from any dataset in any disciplinary. The scalability of this framework to the real-world engineering dataset which is sparse and noisy can be verified in Appendix \ref{app:prac}.}

\subsection{\label{sec:regression} Physics-aware ROM}

\textcolor{black}{To summarize the results so far, physics-aware LVs are extracted by estimating the independence and information intensity of each LVs. These physics-aware LVs are strongly correlated to physical parameters which are the generating factor of training dataset, significantly increasing the interpretability. Since LVs act as intermediaries in the prediction process of ROM, their impact on the ROM is bound to be enormous. In this regard, this section investigates the validity and efficiency of physics-aware ROM, which utilizes physics-aware LVs for the prediction of high-dimensional data.}

\begin{figure}[htb!]
    \centering
	    \includegraphics[width=.6\textwidth]{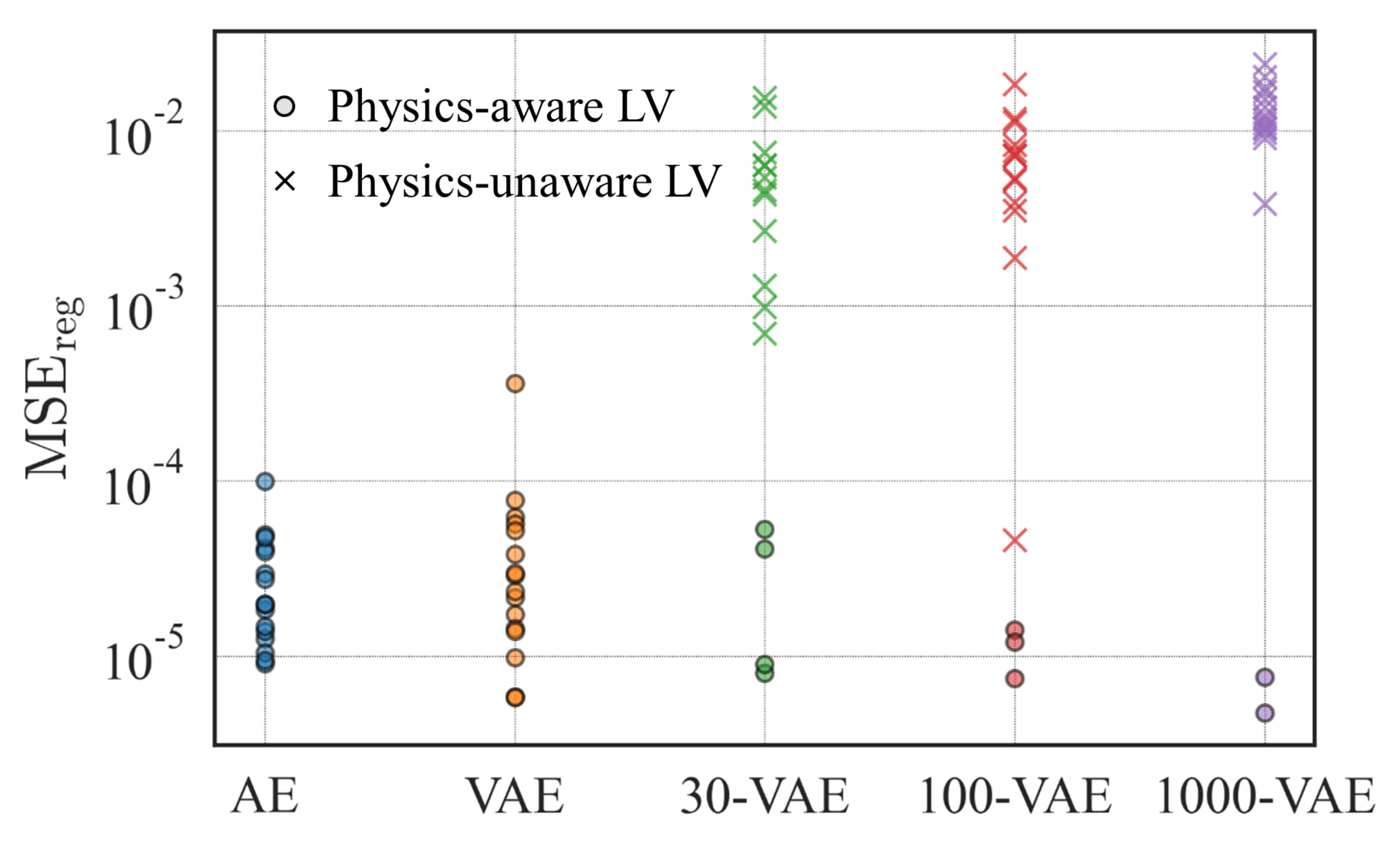}
	\caption{MSE of the regression models in ROM.}
	\label{fig:/MSE_reg_active}
\end{figure}

\begin{figure}[htbp!]
	\centering
	\subfloat[\label{fig:RS_1000VAE_D}]{\includegraphics[width=0.4\textwidth]{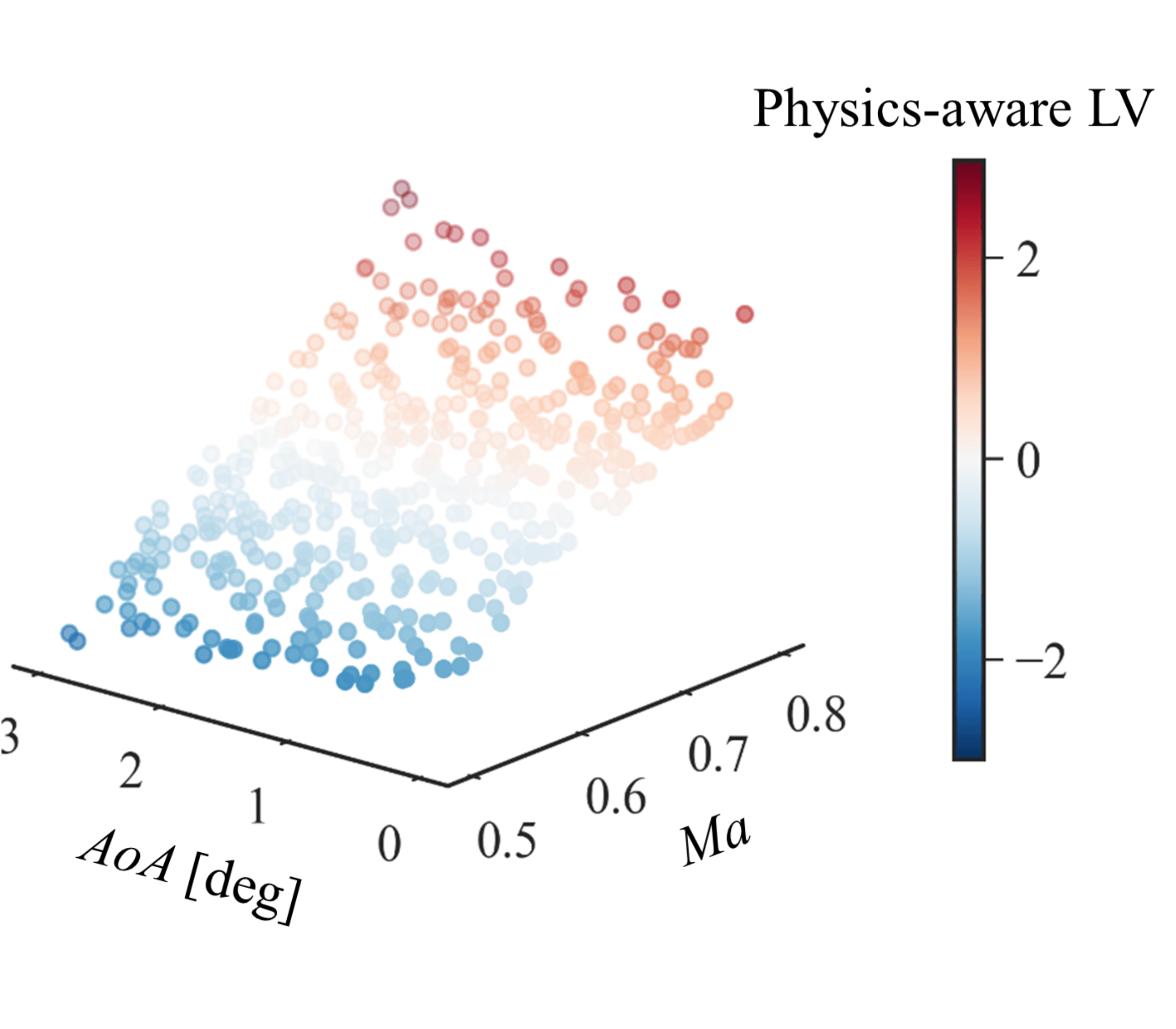}}
	\hfill
    \subfloat[\label{fig:RS_1000VAE_T}]{\includegraphics[width=0.4\textwidth]{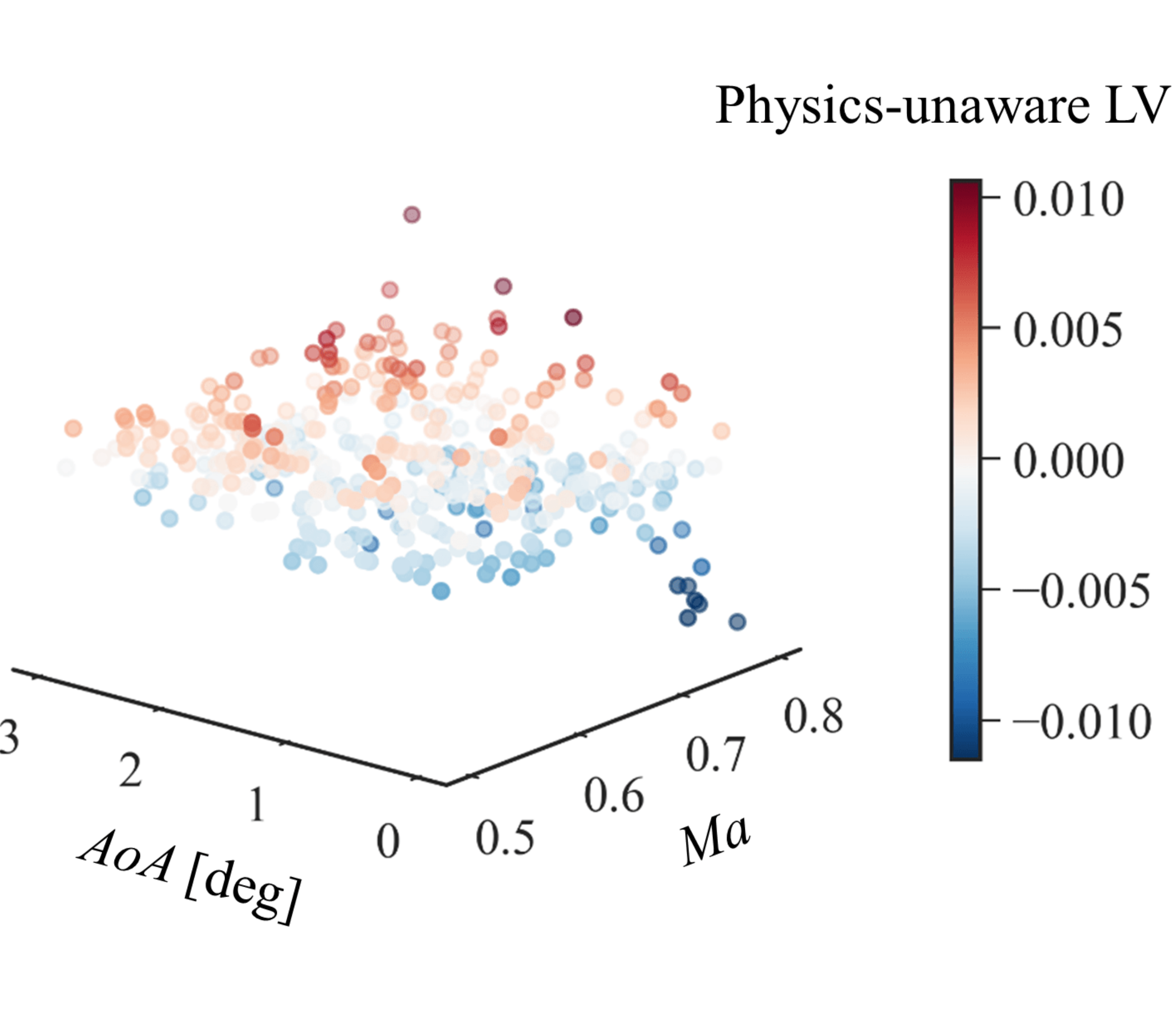}}

    \caption{Comparison of the distribution of two LVs in the 1000-VAE: (a) physics-aware LV, (b) physics-unaware LV.}
     \label{fig:RS_1000VAE}
\end{figure}

Fig. \ref{fig:/MSE_reg_active} shows the MSE of the regression models in ROM, \textcolor{black}{$\rm MSE_{\rm reg}$; it is calculated between true LV and predicted LV by regression model (please refer to Fig. \ref{fig:ROM_structure} for more details about regression models in ROM).} In that 16 regression models are required due to 16 LVs, each point represents the MSE of each model, and the symbol o/x indicates whether each LV is physics-aware or physics-unaware. The results show that the \textcolor{black}{$\rm MSE_\mathrm{reg}$} values of physics-unaware LVs are considerably higher than those of physics-aware LVs. It is because the correlations between physical parameters and physics-unaware LVs are too trivial to be trained by regression models. Fig. \ref{fig:RS_1000VAE} supports this; the distribution of physics-unaware LV (Fig. \ref{fig:RS_1000VAE_T}) is much noisier than that of the physics-aware LV (Fig. \ref{fig:RS_1000VAE_D}). Another interesting point in Fig. \ref{fig:/MSE_reg_active} is that \textcolor{black}{$\rm MSE_\mathrm{reg}$} values of the physics-aware LVs decrease as $\beta$ increases. This implies that the tight coupling between physics-aware LVs and physical parameters is advantageous in terms of regression performance in the ROM process.

\begin{figure}[htb!]
    \centering
    \includegraphics[width=0.6\textwidth]{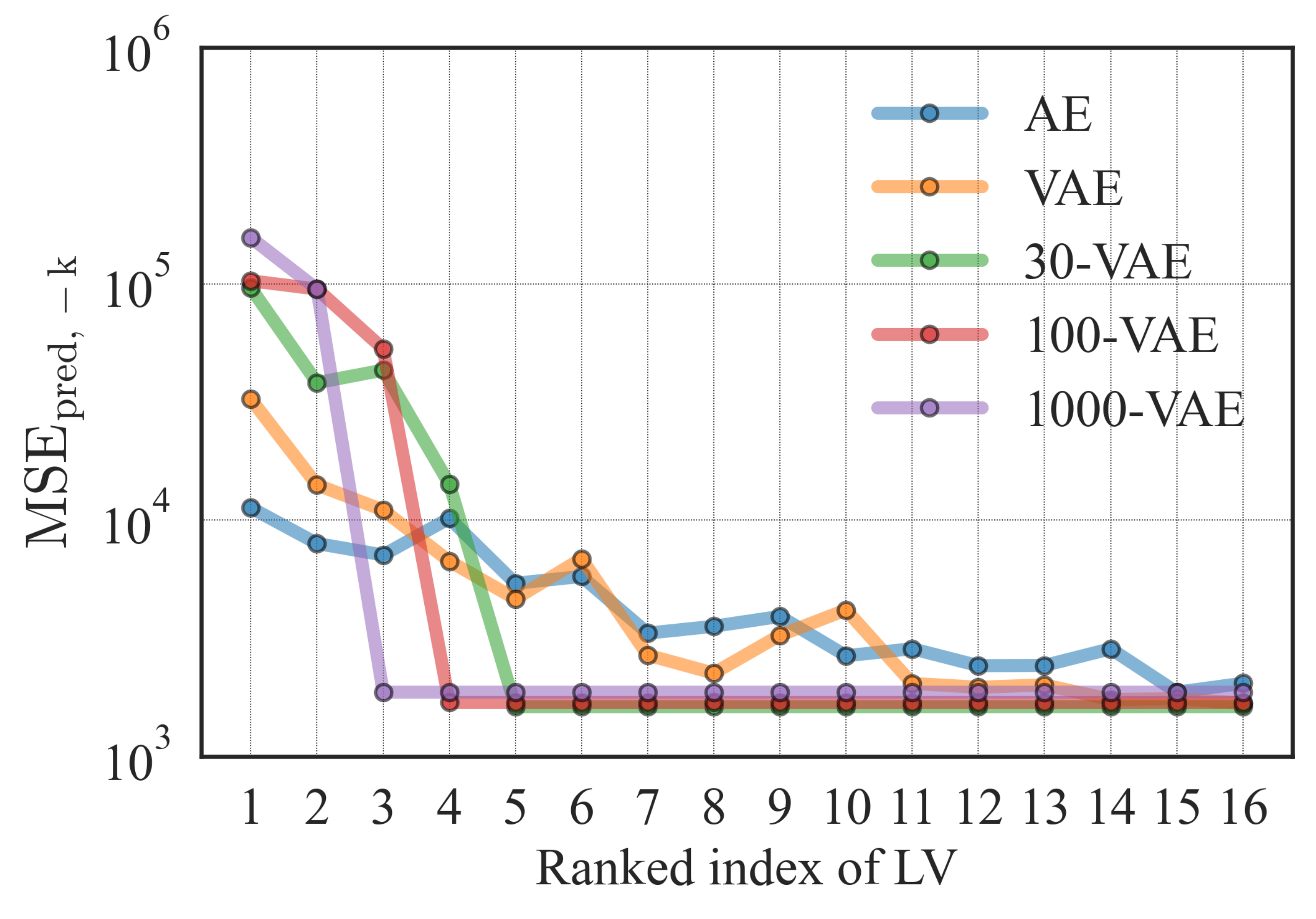}
    \caption{MSE of ROM prediction with the exclusion of $\rm k^{th}$ LV.}
    \label{fig:/MSE_pred_loo}
\end{figure}

\begin{figure}[htb!]
    \centering
    \includegraphics[width=0.65\textwidth]{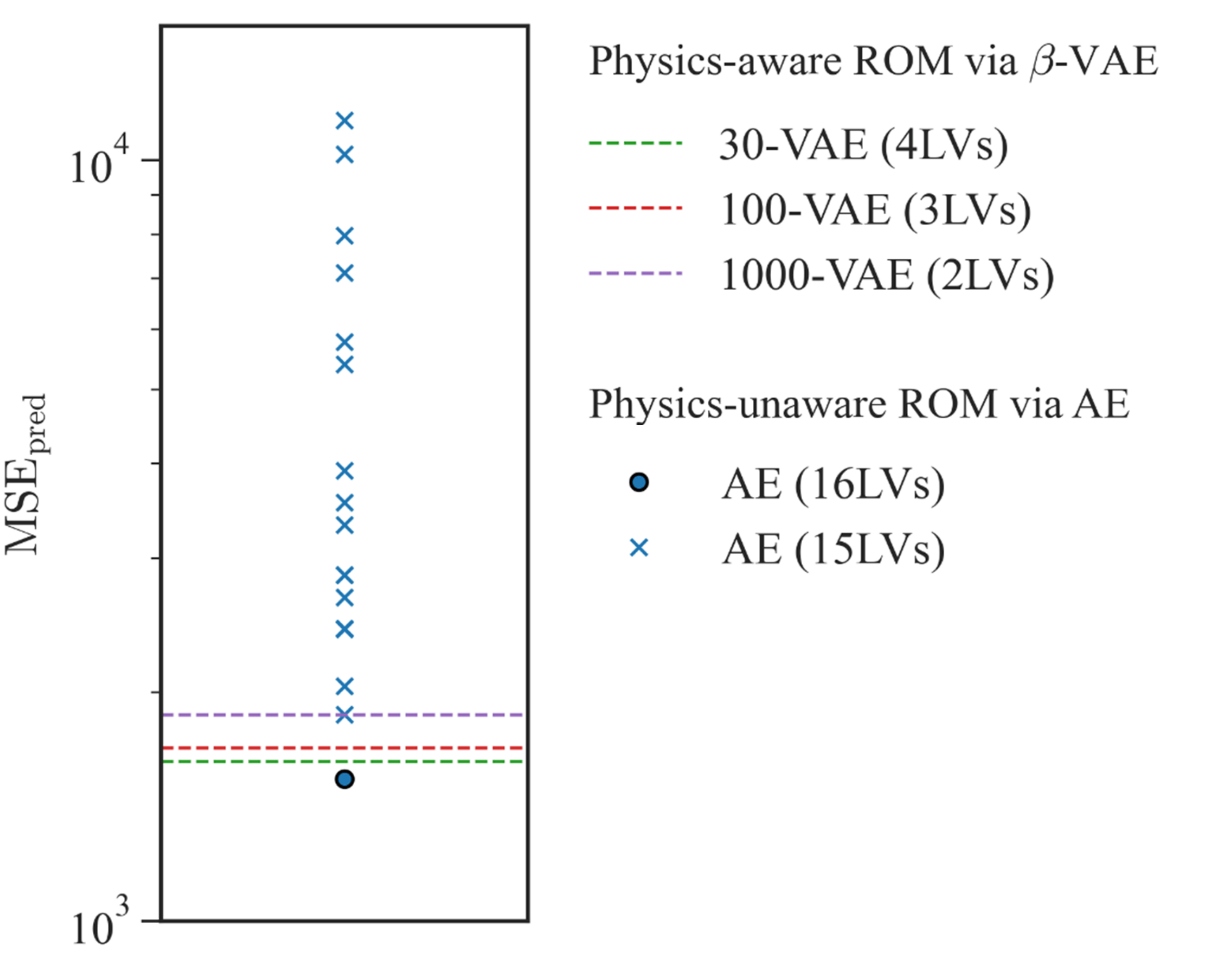}
    \caption{Comparison of prediction MSE between physics-aware ROM and physics-unaware ROM.}
    \label{fig:/MSE_pred_model_compa}
\end{figure}

\begin{figure}[htbp!]
	\centering
	\subfloat[\label{fig:model_contour_compa_pred}]{\includegraphics[width=1\textwidth]{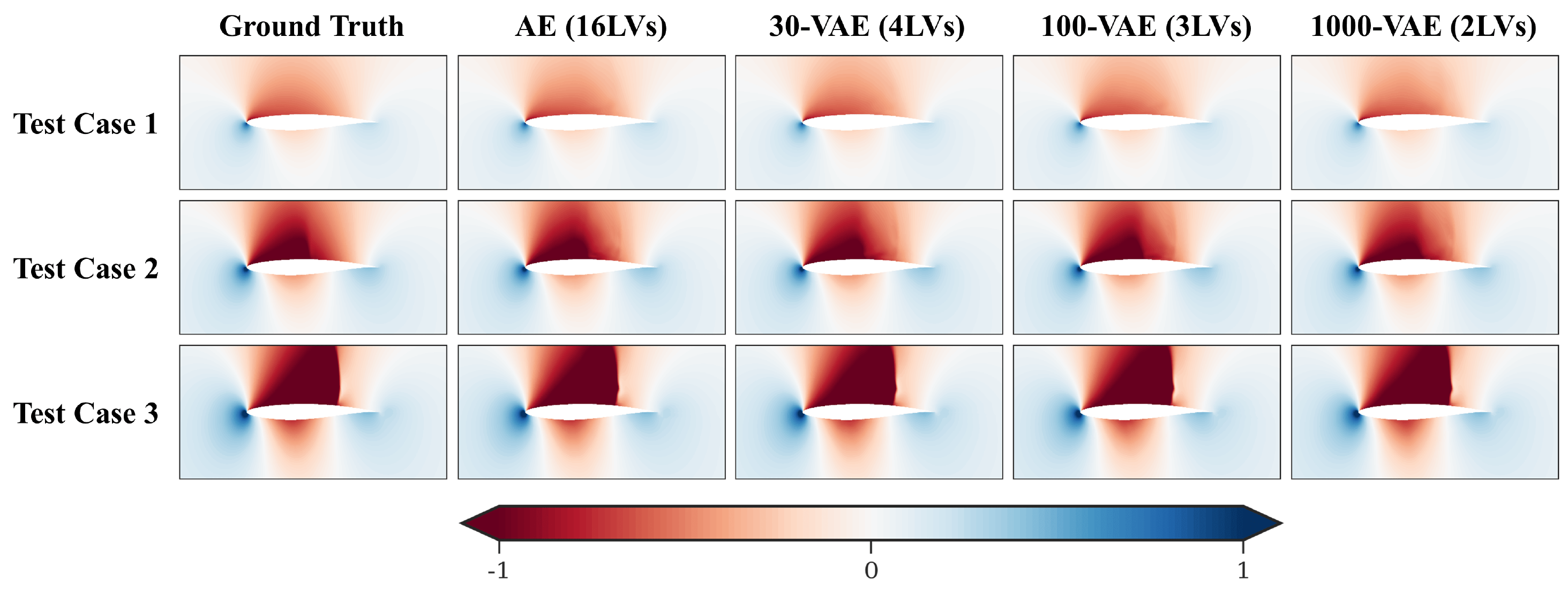}}
	\vfill
    \subfloat[\label{fig:model_contour_compa_error}]{\includegraphics[width=1\textwidth]{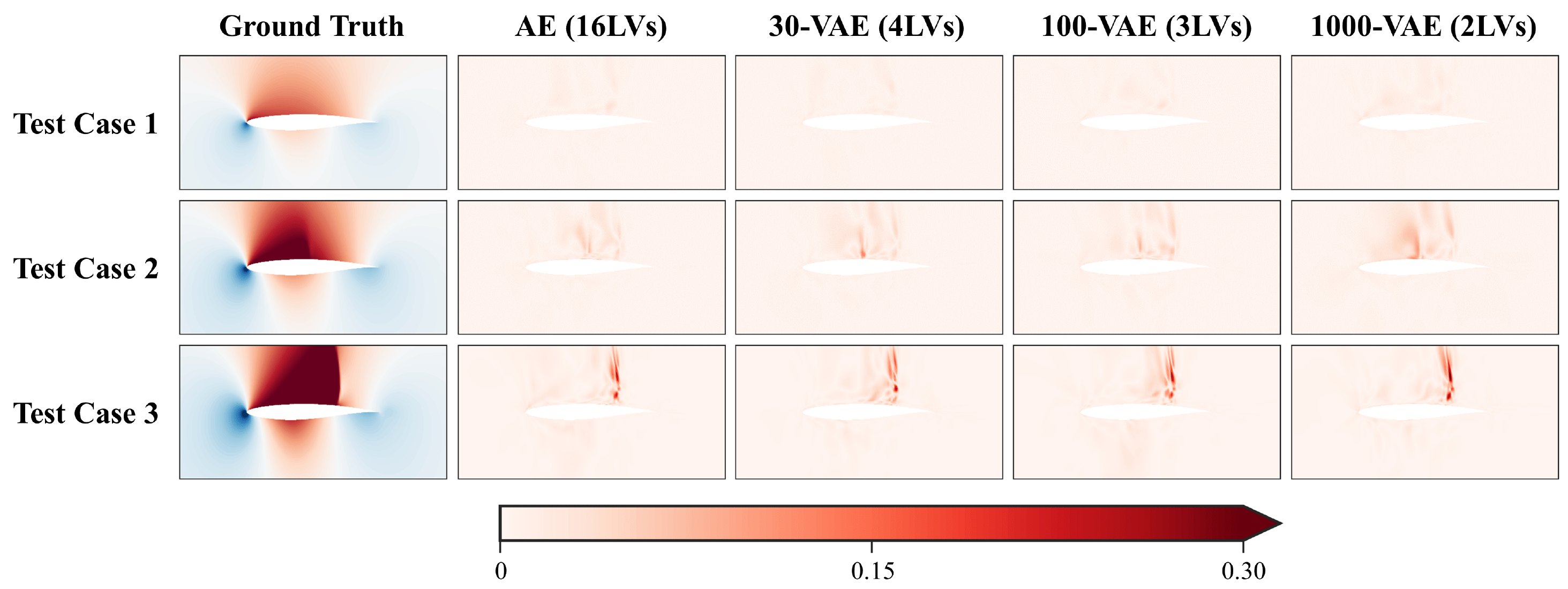}}

    \caption{Pressure contour predicted from AE/$\beta$-VAE-based ROMs: (a) prediction, (b) absolute error.}
     \label{fig:model_contour_compa}
\end{figure}

\textcolor{black}{Then, the MSE between the ground truth flow fields and those predicted by ROM with the exclusion of $\rm k^\mathrm{th}$ LV is calculated, which is denoted as $\rm MSE_{\mathrm{pred, -k}}$ (please note that prediction by ROM means obtaining the flow field from unknown input parameters, as already described in Fig. \ref{fig:ROM_structure}).}  Specifically, $\rm k^\mathrm{th}$ LV is assumed to be constant as $\hat{\mu}_{k}$, whereas the other LVs are predicted from regression models so that the importance of $\rm k^\mathrm{th}$ LV can be investigated. Fig. \ref{fig:/MSE_pred_loo} demonstrates their results, where the x-axis indicates KL-divergence ranking of LVs. An important observation is that the effect of the LV on \textcolor{black}{$\rm MSE_{\mathrm{pred}, -k}$} decreases as the LV ranks down. Likewise, the physics-unaware LVs, those ranked after $\rm 4^\mathrm{th}$ in 30-VAE, after $\rm 3^\mathrm{rd}$ in 100-VAE, and after $\rm 2^\mathrm{nd}$ in 1000-VAE, have negligible effects on \textcolor{black}{$\rm MSE_{\mathrm{pred}, -k}$}: from the fact that the LV ranked higher by KL-divergence has a greater effect on ROM accuracy, it can be concluded that the proposed ranking approach is also valid in terms of ROM performance. \textcolor{black}{This implies that training regression models of physics-unaware LVs are meaningless with respect to the prediction accuracy of ROM. In other words, it is sufficient enough to train regression models of only physics-aware LVs.} In this regard, to examine the necessity of each LV in ROM prediction, Fig. \ref{fig:/MSE_pred_model_compa} visualizes the $\rm MSE_\mathrm{pred}$ values of physics-aware ROM via $\beta$-VAE, where all physics-unaware LVs are excluded (e.g., only two LVs are used for ROM via 1000-VAE). For the comparison, those of physics-unaware ROM via AE are also shown: herein, $\rm MSE_\mathrm{pred}$ utilizing all 16 LVs or 15 LVs with one LV excluded is presented. Although a significantly small number of LVs are used in physics-aware ROM, its accuracy is comparable to that of AE with 16LVs, physics-unaware ROM. \textcolor{black}{Furthermore, even if only one LV is excluded in AE, their ROM accuracy becomes equivalent to or even lower than that of 1000-VAE only with two LVs. This result highlights the inefficiency of ROM through AE, in that it requires all 16 entangled LVs and therefore requires training 16 regression models. Using physics-aware ROM via 30-VAE, the equivalent accuracy can be achieved with only 4 regression models. To visually confirm these results, Fig. \ref{fig:model_contour_compa} shows the pressure contour of physics-aware ROM via $\beta$-VAE and physics-unaware ROM via AE. Herein, the indiscernible difference between them can be confirmed, indicating their equivalent prediction accuracy.}

\clearpage 
\section{\label{sec:conclustion} Conclusion}
\textcolor{black}{This study proposed the physics-aware ROM based on physics-aware LVs, which are interpretable and information-intensive LVs extracted by $\beta$-VAE. The proposed framework is validated with the 2D transonic benchmark problem in the following order. First, the extraction process of physics-aware LVs is scrutinized by quantitatively estimating their independence and information intensity. Then, the actual physical meanings of these LVs are thoroughly investigated. Finally, the effectiveness of the proposed physics-aware ROM compared to conventional ROMs are verified. The key contributions of our study can be summarized as follows:}

\begin{enumerate}

    \item \textcolor{black}{
    The impacts of hyperparameter $\beta$ on the independence of LVs were scrutinized, and its effect on the independence of LVs was practically confirmed in that LVs become disentangled from each other as $\beta$ increases.}
  
    \item \textcolor{black}{KL-divergence ranking method was proposed to measure the information intensity of each LV. This approach has two following advantages over the previous ranking method: KL-divergence is the direct cause of the discrepancies in information intensity of LVs and it does not require cumbersome post-processing of reconstructed data. The proposed criterion was confirmed to have a consistent trend with estimated standard deviations and Sobol indices, indicating their validity. Through this ranking method, the effect of $\beta$ on the latent space regularization was practically confirmed in that LVs become information-intensive as $\beta$ increases.}
    
    \item \textcolor{black}{The physical meanings contained in physics-aware LVs were thoroughly investigated. The correlation between the physical generating factors of the training dataset and the information physics-aware LVs contain was scrutinized as $\beta$ varies. Finally, it was confirmed quantitatively and qualitatively that the extracted physics-aware LVs in 1000-VAE actually correspond to the generating factors, which were $Ma$ and $AoA$ in this study. To the best of the authors' knowledge, this is the first study to practically confirm that $\beta$-VAE can automatically extract the physical generating factors in the field of applied physics.}

    \item \textcolor{black}{The effects of physics-awareness of LVs on the accuracy of regression and prediction processes in ROM are analyzed and it was confirmed that only physics-aware LVs had a significant effect on their accuracy. Therefore, physics-aware ROM, which utilizes only physics-aware LVs is proposed for its efficiency in that the number of required regression models can be reduced significantly. Finally,  compared to the conventional ROMs, its validity and efficiency were successfully verified.}

\end{enumerate}

\textcolor{black}{The presented data-driven physics-aware ROM has great potential in two engineering applications. First, extraction process of physics-aware LVs can be applied to discovering generating factors from the given dataset. For example, this application can be extended to identify fault-causing factors in sensor data from manufacturing processes. Second, physics-aware ROM can be an efficient alternative to conventional black-box ROMs in that it utilizes necessary regression models of only physically interpretable and information-intensive LVs, rather than redundant regression models of all uninterpretable LVs. Though the application of this framework was demonstrated via 2D transonic benchmark problem, it can be easily applied and extended to numerous engineering disciplines in that no special assumptions have been made on this specific problem. For future work, a more comprehensive investigation on physics-aware LVs will be conducted, such as their scalability to the temporal dataset or their ability to discern redundant physical parameters.}

\section*{Acknowledgement}
This work was supported by a National Research Foundation of
Korea grant funded by the Ministry of Science and ICT (NRF-2017R1A5A1015311).

\section*{Author Declarations}
\subsection*{Author Contributions}
Yu-Eop Kang and Sunwoong Yang contributed equally to this work.

\subsection*{Conflict of Interest}
The authors declare that they have no known competing financial interests or personal relationships that could have appeared to influence the work reported in this paper.

\section*{Data availability}
The data that support the findings of this study are available from the corresponding author upon reasonable request.

\appendix
\section{\label{app:POD} POD-based ROM results}

\begin{figure}[htbp!]
	\centering
	\subfloat[\label{fig:LT_POD_1d_1LV}]{\includegraphics[width=0.45\textwidth]{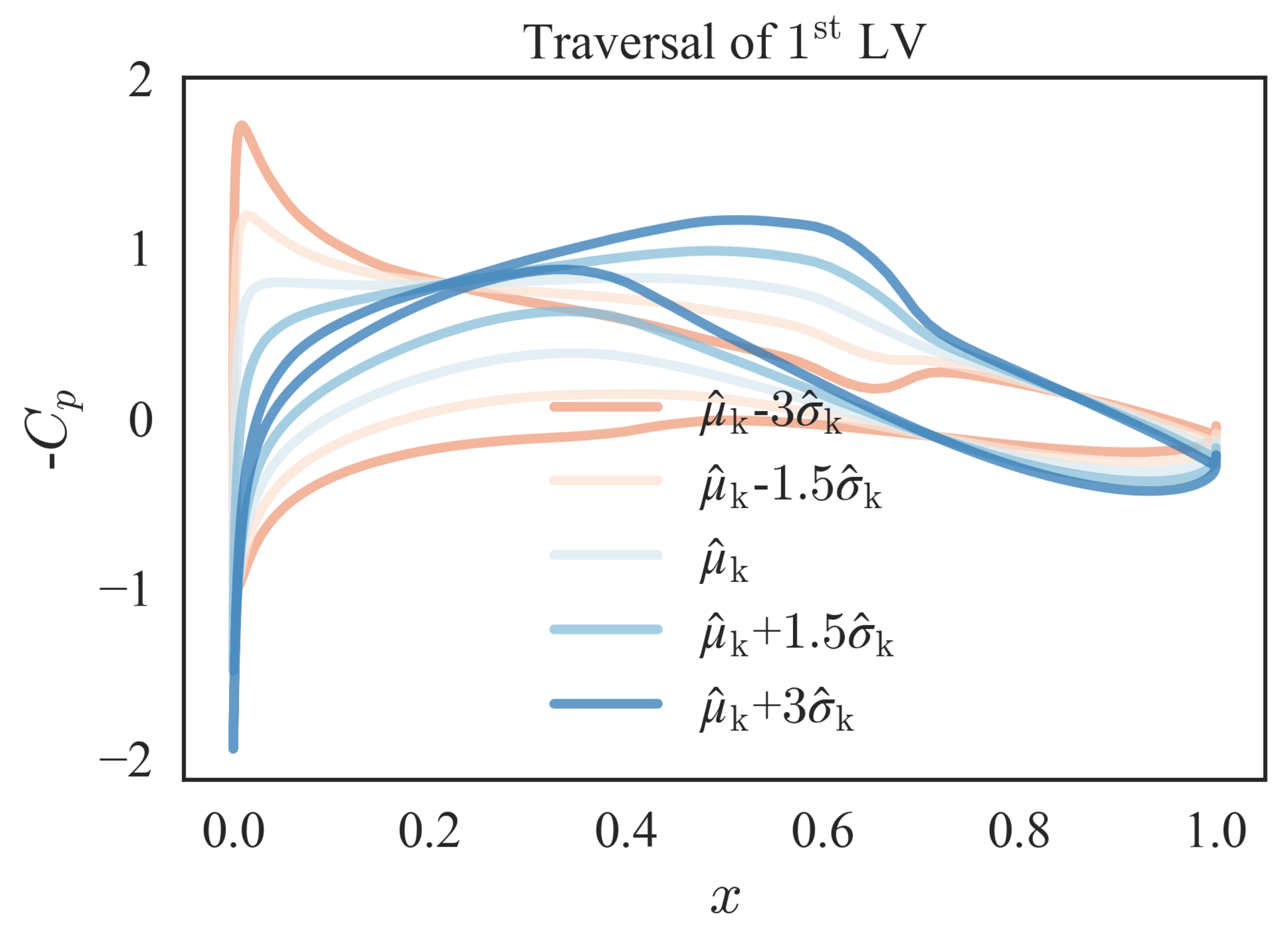}}
	\hfill
    \subfloat[\label{fig:LT_POD_1d_2LV}]{\includegraphics[width=0.45\textwidth]{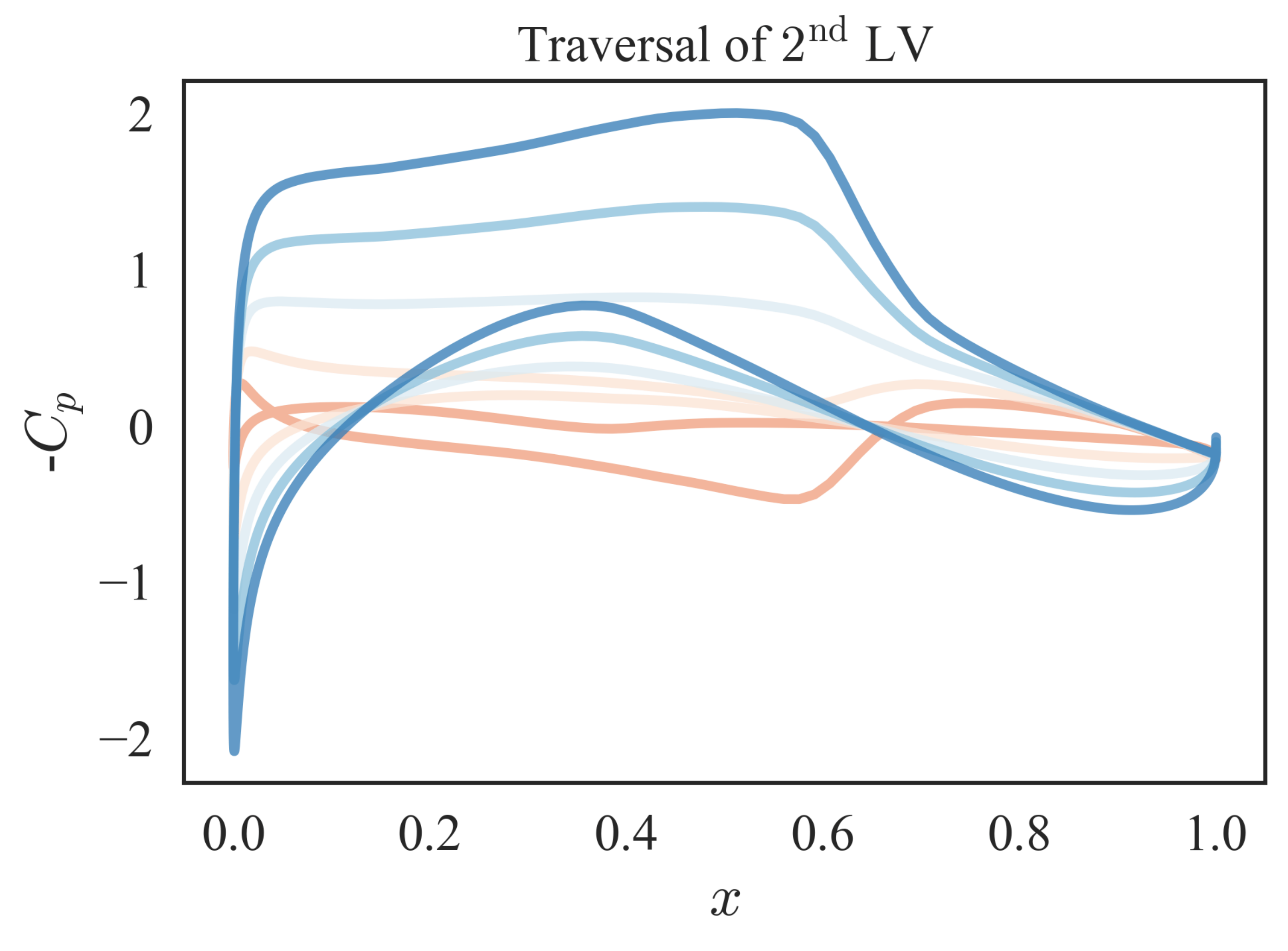}}

    \caption{Latent traversal plots of airfoil surface pressure distributions in POD : (a) traversal of $\rm 1^{st}$ LV, and (b) traversal of $\rm 2^{nd}$ LV.}
     \label{fig:LT_POD_1d}
\end{figure}

\textcolor{black}{While the excellence of AE-based ROM over POD-based ROM has already been conducted by numerous studies \cite{milano2002neural, eivazi2022towards, zhang2021machine}, it is worth reaffirming from the perspective of this study. Therefore, additional results by POD-based ROM are presented in this section. Fig. \ref{fig:LT_POD_1d} shows the same results as in Fig. \ref{fig:cp_ma_aoa} except it is based on POD. When compared to 1000-VAE in Fig. \ref{fig:cp_ma_aoa}, the discontinuity due to the shock wave is not sharply captured by the top two dominant LVs of POD. Moreover, these two LVs encode two physical characteristics, the presence of the shock wave and the magnitude of the suction peak, in an entangled manner, whereas 1000-VAE successfully separated their information in a disentangled manner. To quantitatively verify the entanglement of POD LVs, the same analysis using LR models with two LVs (as Eq. \ref{eq:bi_lr}) is repeated, and the results are as follows:}

\begin{equation}\
\label{eq:bi_lr_pod}
\begin{bmatrix}
Ma \\
AoA 
\end{bmatrix} = 
\begin{bmatrix}
0.448 & 0.862\\
-0.897 & 0.433
\end{bmatrix}
\begin{bmatrix}
1^{st}\; LV \\
2^{nd}\; LV
\end{bmatrix}.
\end{equation}
\textcolor{black}{As can be confirmed, both LVs contribute significantly to the prediction of each physical parameter (herein, test dataset $R^2$ values for Ma and AoA are calculated as 0.949 and 0.994), meaning that they cannot solely represent the $Ma$ and $AoA$ as 1000-VAE model. All these results indicate that despite the POD algorithm ensuring orthogonality, the physical interpretability of its latent space cannot be guaranteed.}

\begin{figure}[htbp!]
	\centering
	\subfloat[\label{fig:Model_compa_POD_a}]{\includegraphics[width=0.645\textwidth]{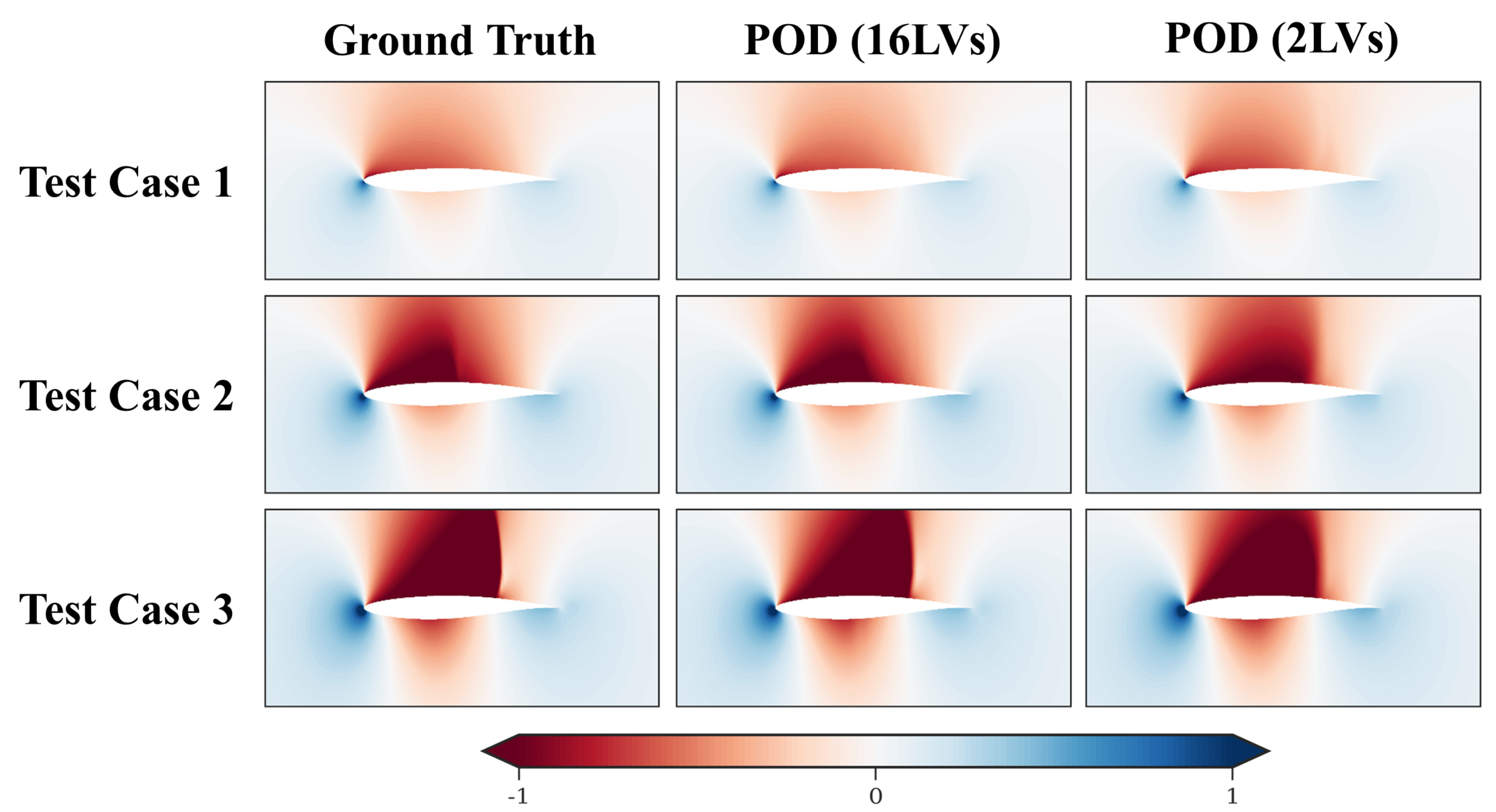}}
	\hfill
    \subfloat[\label{fig:Model_compa_POD_b}]{\includegraphics[width=0.355\textwidth]{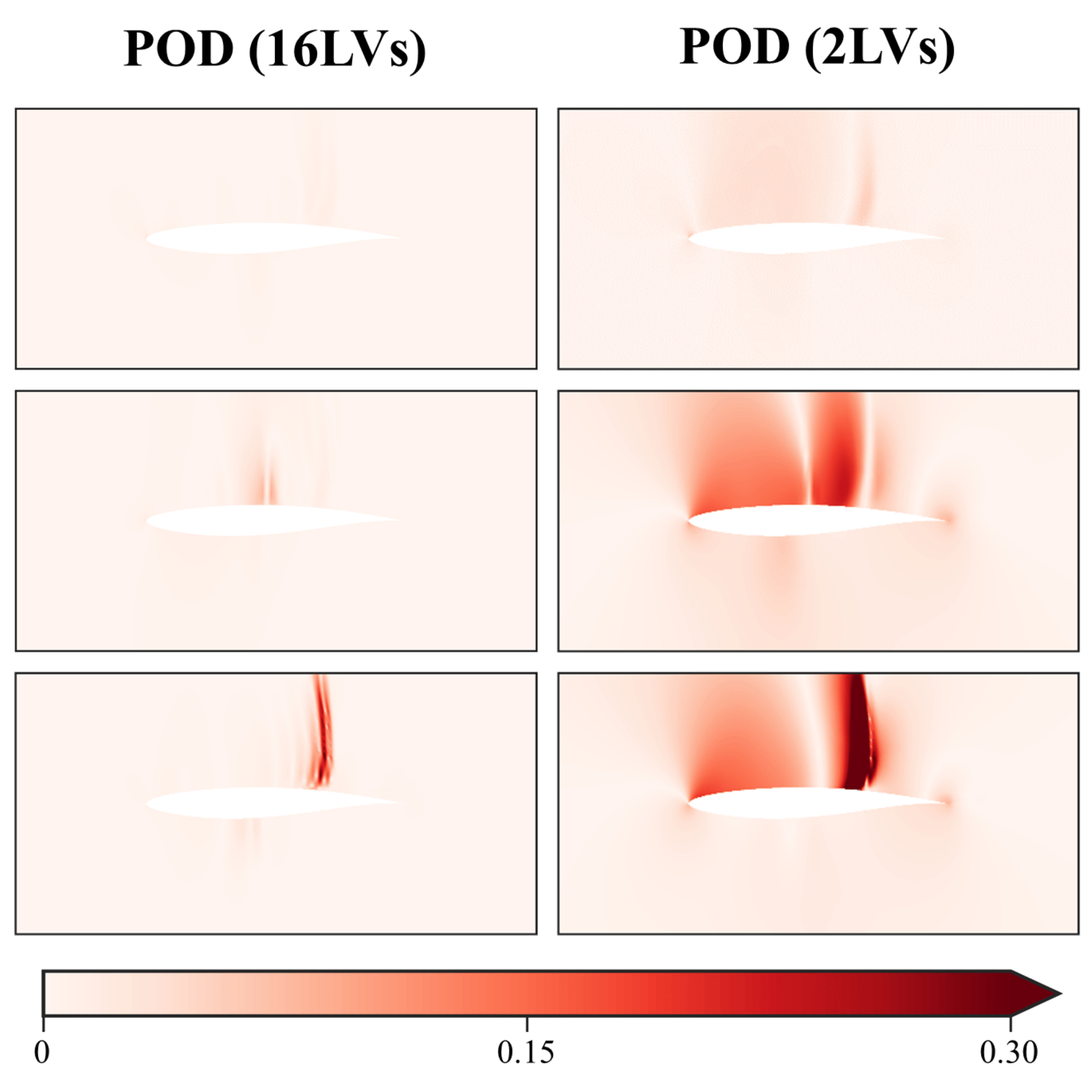}}

    \caption{Pressure contour predicted from POD-based ROM: (a) prediction, (b) absolute error.}
     \label{fig:Model_compa_POD}
\end{figure}

\textcolor{black}{To inspect the prediction accuracy of POD-based ROM, Fig. \ref{fig:Model_compa_POD} shows the pressure contour predicted from POD-based ROM and its absolute error contour. Herein, POD (16LVs) and POD (2LVs) each represents ROM with 16 and 2 LVs (or modes) of POD. When compared to Fig. \ref{fig:model_contour_compa}, the error contour of POD with 16LVs is comparable to that of AE/VAE/$\beta$-VAE. However, when the number of LVs is reduced to 2, which is the number of generating factors of training dataset, the error contour is much more worse. Indeed, $\rm MSE_\mathrm{pred}$ of ROM based on POD with 2LVs ($4.94\times 10^4$) is 26 times higher than that of 1000-VAE with 2LVs ($1.87\times 10^3$). More interestingly, even that of POD with 16LVs ($2.07\times10^3$) is 1.1 times higher than 1000-VAE with 2LVs. Despite the further increment of the number of LVs used in POD-based ROM to 64, it is confirmed that the $\rm MSE_\mathrm{pred}$ of POD ($1.93\times 10^3$) is still higher than that of 1000-VAE. These results confirm that POD-based ROM requires an excessive number of LVs compared to $\beta$-VAE-based ROM for ensuring the same prediction accuracy.}

\textcolor{black}{This section shows the obvious superiority of nonlinear-based DR methods over linear-based DR method (POD) in terms of physical interpretability of latent space and consequent prediction accuracy of ROM, both of which are key factors of this study.}

\section{\label{app:prac} Scalability of the framework for extracting physics-aware LVs}

\begin{figure}[htbp!]
	\centering
	\subfloat[\label{fig:Noise_Cp}]{\includegraphics[width=0.47\textwidth]{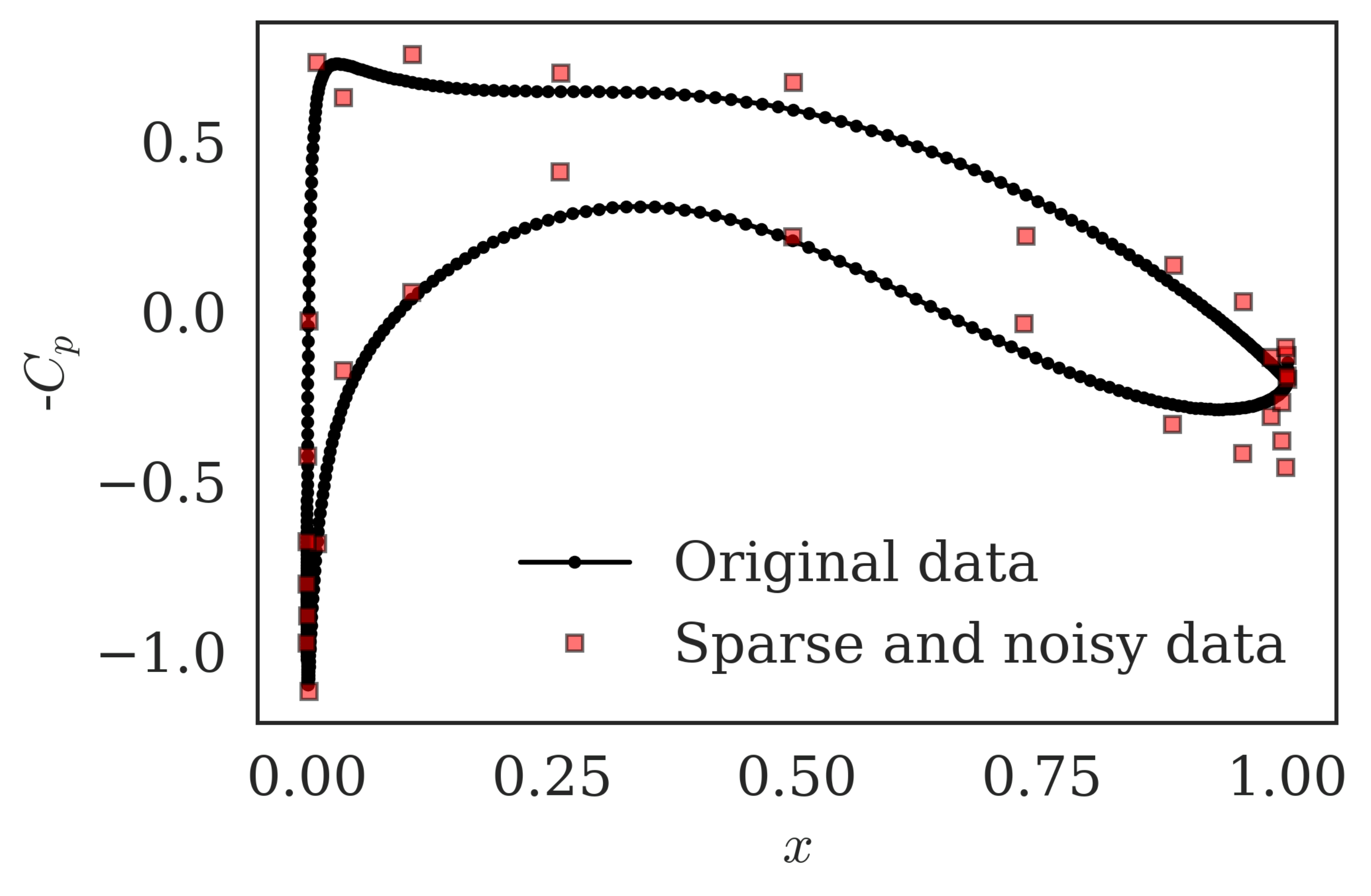}}
	\hfill
    \subfloat[\label{fig:Noise_CL}]{\includegraphics[width=0.44\textwidth]{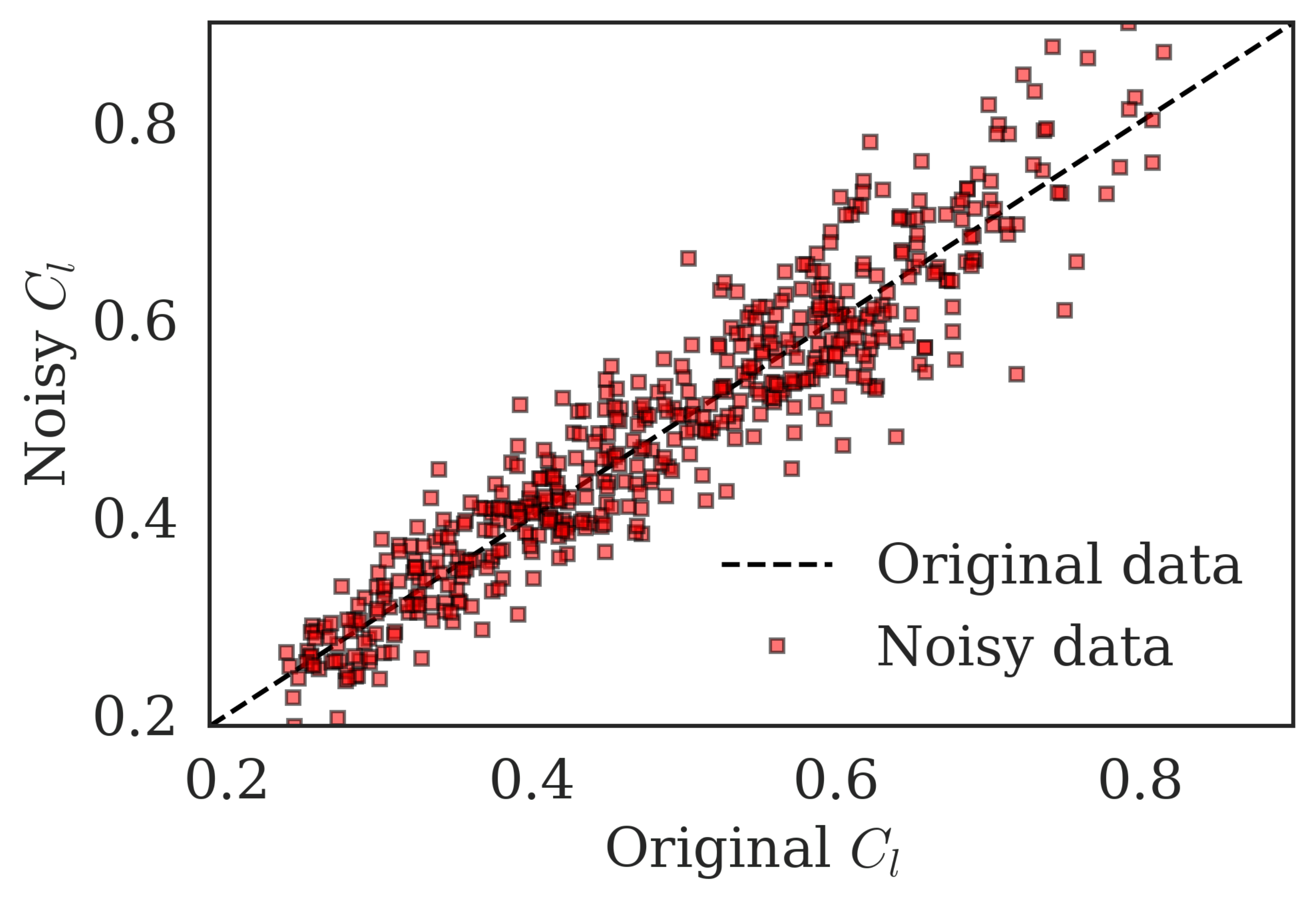}}
    \vfill
    \subfloat[\label{fig:Noise_CD}]{\includegraphics[width=0.44\textwidth]{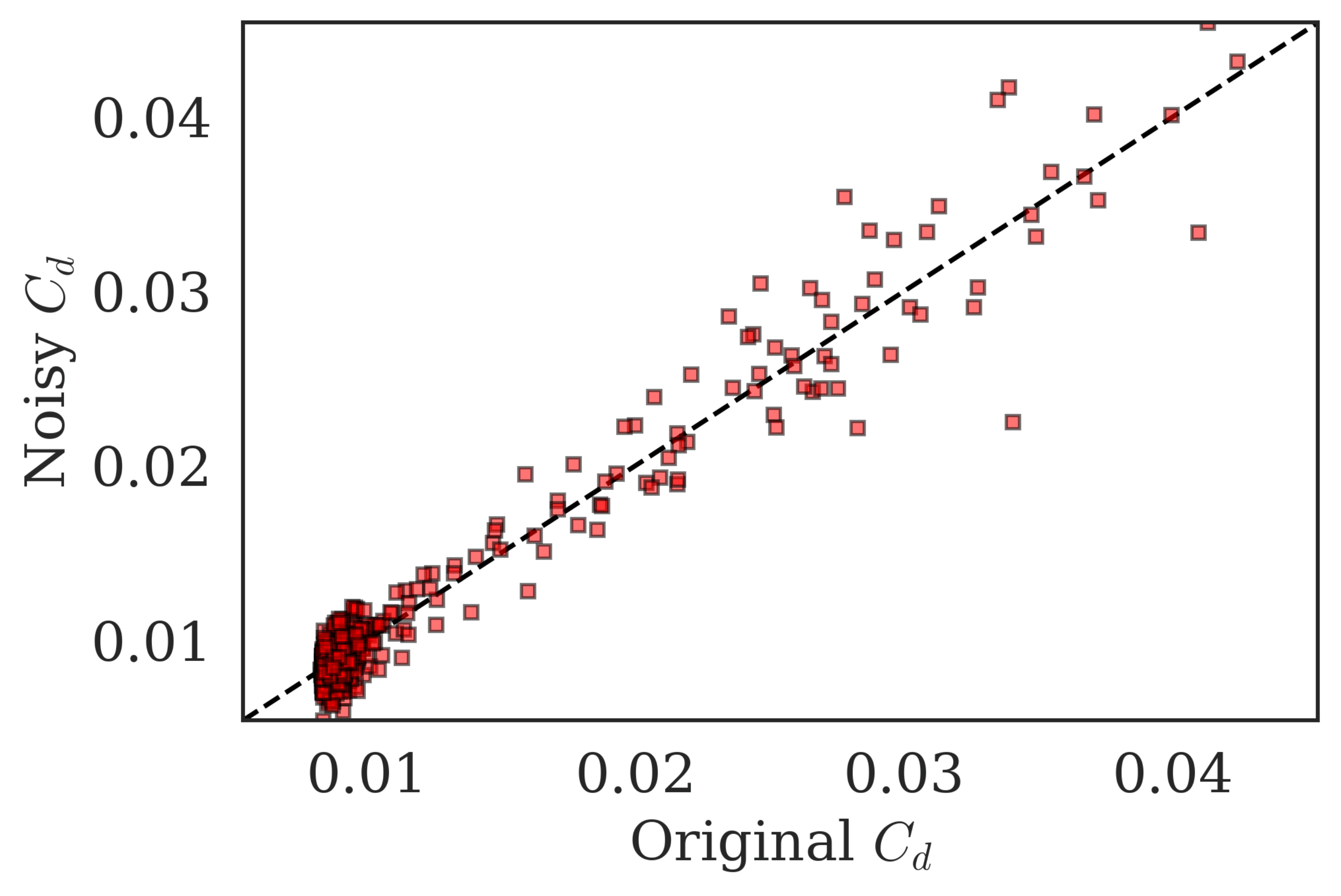}}
    \hfill
    \subfloat[\label{fig:Noise_CM}]{\includegraphics[width=0.46\textwidth]{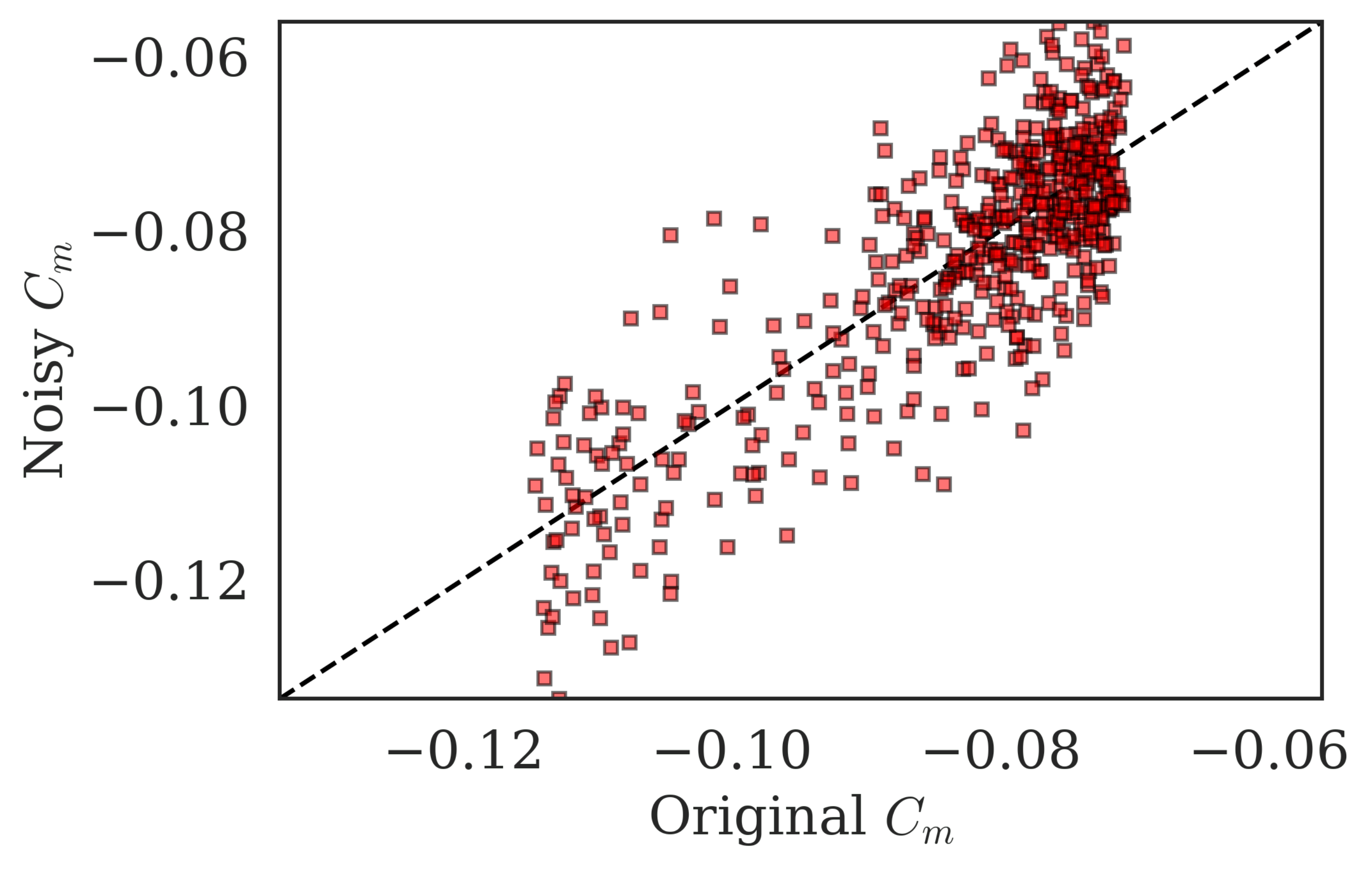}}
    
    \caption{Training dataset before and after preprocessing: (a) 32 surface pressure values, (b) $C_l$, (c) $C_d$, and (d) $C_m$. One training sample consists of all 32 values in (a) and one value each in (b), (c), and (d).}
    \label{fig:Noise_coeff}
\end{figure}

\begin{figure}[htb!]
    \centering
    \includegraphics[width=0.8\textwidth]{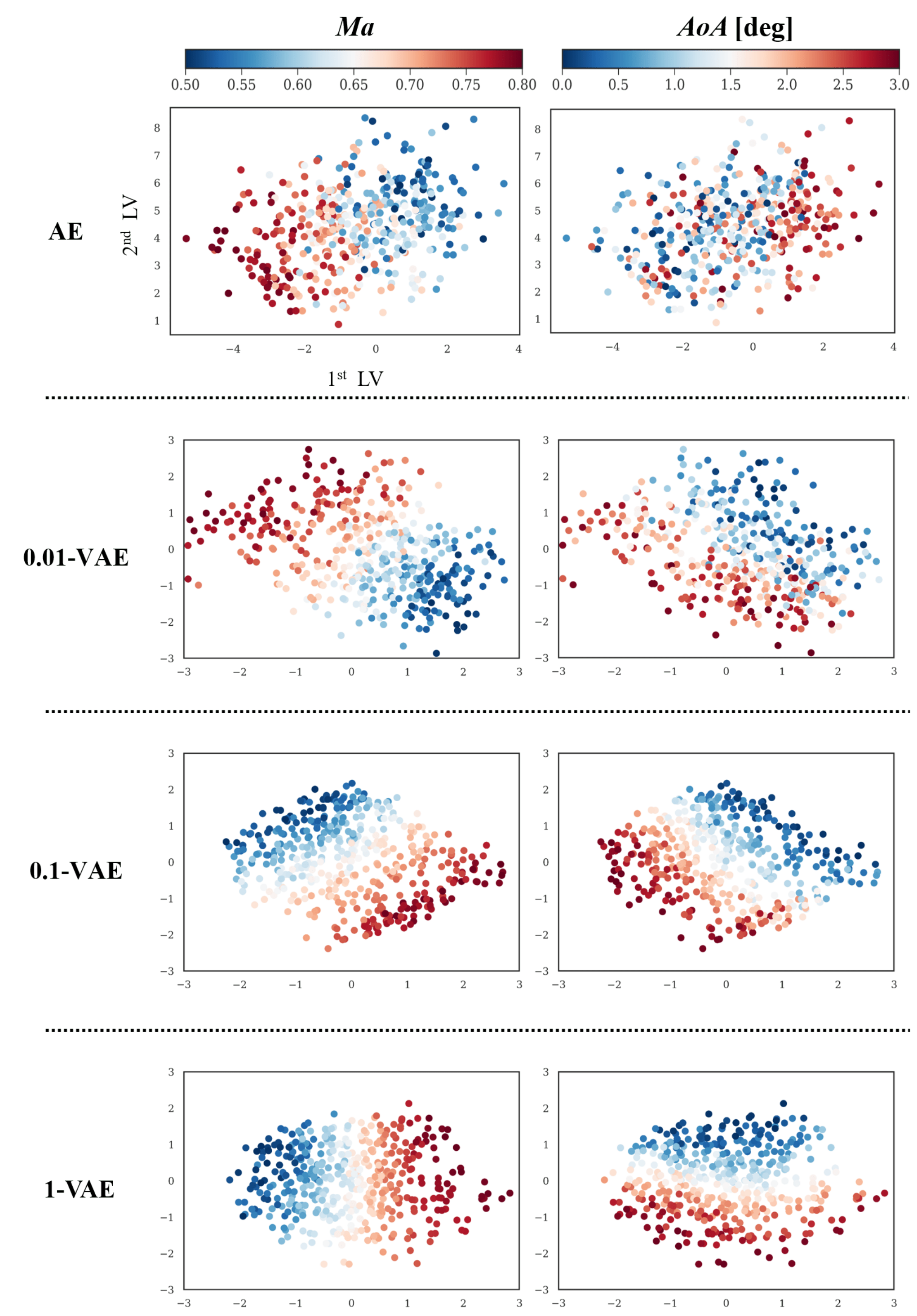}
    \caption{Investigation of physical features contained in the top two LVs for sparse and noisy datasets.}
    \label{fig:/add_exp_latent_vis}
\end{figure}

\textcolor{black}{In the main text, the framework for extracting physics-aware LVs using $\beta$-VAE is validated with regularized and well-organized CFD dataset. However, most of the data in real-world engineering is sparse and noisy. Accordingly, in this section, the practical scalability of the proposed framework is verified by utilizing only a small portion of the training dataset and adding artificial noises. For this purpose, the training dataset in the main text (which is the tensor of $3 \times 512 \times 128$) is preprocessed to be a vector with only 35 elements: 32 surface pressure values, lift coefficient ($C_{l}$), drag coefficient ($C_{d}$), and pitching moment coefficient ($C_{m}$). Artificial Gaussian noises are added considering the scale of each element, and the final dataset are shown in Fig. \ref{fig:Noise_coeff}. Then, AE and $\beta$-VAE models ($\beta\in[0.01, 0.1, 1]$) are trained and their top two ranked LVs are visualized in Fig. \ref{fig:/add_exp_latent_vis} (which corresponds to Fig. \ref{fig:ma_aoa} in Sec. \ref{sec:Disen}). Again, it can be seen that LVs of AE are highly entangled. On the other hand, LVs of $\beta$-VAE become more and more disentangled as $\beta$ increases so that eventually in 1-VAE, each LV solely represents $Ma$ and $AoA$, respectively. It can be concluded that the extraction of physics-aware LVs via $\beta$-VAE, which is first observed and reported in this study, has the potential to be applied to real-world engineering problems where training datasets are sparse and noisy.}

\clearpage
\nocite{*}
\bibliography{references}
\bibliographystyle{unsrt}

\end{document}